\documentclass[review,12pt,authoryear]{elsarticle}

\usepackage{amssymb}
\usepackage{amsthm}
\usepackage{amsmath}
\usepackage{mathrsfs}
\usepackage{graphicx}
\usepackage{epstopdf}
\usepackage{float}
\usepackage{caption}
\usepackage{subcaption}
\usepackage{bm}
\usepackage{amssymb} 
\usepackage{extarrows} 
\usepackage{bbm}
\usepackage{mathrsfs}
\usepackage{hyperref} 
\usepackage{cleveref}
\usepackage{soul}
\usepackage{booktabs}
\usepackage{tabu} 
\usepackage{longtable}
\usepackage{xcolor} 
\usepackage[margin=2.2cm]{geometry}
\usepackage{enumitem} 

\newcommand{\Hsquare}{
  \text{\fboxsep=-.2pt\fbox{\rule{0pt}{1ex}\rule{1ex}{0pt}}}%
}

\journal{Journal of the Mechanics and Physics of Solids}

\makeatletter
\def\@author#1{\g@addto@macro\elsauthors{\normalsize%
    \def\baselinestretch{1}%
    \upshape\authorsep#1\unskip\textsuperscript{%
      \ifx\@fnmark\@empty\else\unskip\sep\@fnmark\let\sep=,\fi
      \ifx\@corref\@empty\else\unskip\sep\@corref\let\sep=,\fi
      }%
    \def\authorsep{\unskip,\space}%
    \global\let\@fnmark\@empty
    \global\let\@corref\@empty  
    \global\let\sep\@empty}%
    \@eadauthor={#1}
}
\makeatother

\begin{document}

\begin{frontmatter}



\title{A generalised, multi-phase-field theory for dissolution-driven stress corrosion cracking and hydrogen embrittlement}


\author{Chuanjie Cui\fnref{IC,T}}


\author{Rujin Ma \fnref{T,KL}}

\author{Emilio Mart\'{\i}nez-Pa\~neda\corref{cor1}\fnref{IC}}
\ead{e.martinez-paneda@imperial.ac.uk}

\address[IC]{Department of Civil and Environmental Engineering, Imperial College London, London SW7 2AZ, UK}

\address[T]{College of Civil Engineering, Tongji University, Shanghai 200092, China}

\address[KL]{Key Laboratory of Performance Evolution and Control for Engineering Structures, Tongji University, Shanghai 200092, China}

\cortext[cor1]{Corresponding author.}

\begin{abstract}
We present a phase field-based electro-chemo-mechanical formulation for modelling mechanics-enhanced corrosion and hydrogen-assisted cracking in elastic-plastic solids. A multi-phase-field approach is used to present, for the first time, a general framework for stress corrosion cracking, incorporating both anodic dissolution and hydrogen embrittlement mechanisms. We numerically implement our theory using the finite element method and defining as primary kinematic variables the displacement components, the phase field corrosion order parameter, the metal ion concentration, the phase field fracture order parameter and the hydrogen concentration. Representative case studies are addressed to showcase the predictive capabilities of the model in various materials and environments, attaining a promising agreement with benchmark tests and experimental observations. We show that the generalised formulation presented can capture, as a function of the environment, the interplay between anodic dissolution- and hydrogen-driven failure mechanisms; including the transition from one to the other, their synergistic action and their individual occurrence. Such a generalised framework can bring new insight into environment-material interactions and the understanding of stress corrosion cracking, as demonstrated here by providing the first simulation results for Gruhl's seminal experiments.\\
\end{abstract}

\begin{keyword} Multi-phase-field \sep Stress corrosion cracking \sep Anodic dissolution \sep Hydrogen embrittlement  \sep Fracture mechanics



\end{keyword}

\end{frontmatter}

{\footnotesize
\tableofcontents}


\section{Introduction}
\label{Sec:Intro} 


Predicting the failure of engineering components in aggressive environments is a longstanding scientific challenge with important technological implications \citep{Turnbull1993,RILEM2021}. The combination of mechanical load and corrosive environments facilitates the nucleation and growth of cracks, in what is commonly referred to as \textit{stress corrosion cracking} (SCC). The term stress corrosion cracking broadly encapsulates a wide range of mechanisms. These include local metallic dissolution, rupture of the protective passive film, and uptake of pernicious species such as hydrogen, which can be generated as a result of cathodic reactions in acid environments or through hydrolysis reactions during the corrosion process \citep{Raja2011}. SCC phenomena can be generally classified into two categories. The first one refers to fracture events driven by local material dissolution at the crack tip, as a result of anodic corrosion reactions \citep{Scully1975,Scully1980,Ford1990,MacDonald1991}. The second one is governed by the uptake and diffusion of hydrogen within the crystal lattice, resulting in embrittlement \citep{Gangloff2003,Barrera2018,Djukic2019}. These two classes of phenomena have been investigated extensively, but independently from each other. Examples of anodic-driven models include the surface mobility model \citep{Galvele1987}, anodic reaction-induced cleavage \citep{Sieradzki1985} and slip-dissolution \citep{Andresen1988,Parkins1996,Engelhardt1999}; while hydrogen assisted cracking has been rationalised through decohesion-based mechanisms \citep{Gerberich1991,AM2016}, hydrogen-dislocation interactions \citep{Sofronis1995,Lynch2019} and cleavage driven by hydride formation \citep{Lufrano1998a,Shishvan2020}. However, anodic- and hydrogen-driven SCC phenomena often act in concert and, as a result, there is a need for a generalised theory that can encapsulate their interaction and synergistic effects.\\ 

The various hydrogen and dissolution mechanisms involved in SCC are strongly coupled and encompass various physical phenomena. For example, the presence of a passivation film prevents hydrogen ingress in the metal but the film can be ruptured by crack tip straining. Also, pits resulting from localised material dissolution will facilitate hydrogen uptake by changing the local electrolyte pH and by acting as stress concentrators, as hydrogen accumulates in areas of high hydrostatic stress. The need for capturing multiple chemical and mechanical phenomena, and their interplay, is arguably one of the reasons why predicting SCC is considered a longstanding challenge \citep{Turnbull2001,Jivkov2004}. Another important challenge is the interfacial nature of SCC. The shape of the SCC defect plays a significant role in the electrolyte chemistry, the nearby mechanical fields and the hydrogen uptake; in turn, these factors govern local corrosion kinetics and thus the morphology of the aqueous electrolyte-solid metal interface. Resolving the underlying processes requires tracking the evolution of the corrosion front, a well-known computational challenge \citep{Duddu2016,Dekker2021,Chen2021}. Phase field methods have emerged as a compelling approach for modelling coupled interfacial problems. Evolving interfaces can readily be simulated by using an auxiliary \emph{phase field} variable that takes a distinct value in each of the phases (e.g., 0 and 1) and varies smoothly in-between. The interface then evolves based on the solution to the phase field differential equation, a diffuse interface description that brings multiple advantages: (i) it requires no special treatment or presumptions, as the interface equation is solved in the entire domain; (ii) it can readily handle topological changes such as division or merging of interfaces; and (iii) it can be easily combined with equations describing various physical phenomena. The phase field paradigm has been recently extended to model hydrogen embrittlement \citep{CMAME2018,Wu2020b,TAFM2020c} and material dissolution \citep{Mai2016,Nguyen2018,JMPS2021,Ansari2021,Lin2021}. However, a generalised model encompassing these two key elements of SCC, and their interactions, has not been presented yet.\\

In this work, we present a generalised formulation for dissolution-driven SCC and hydrogen embrittlement. A multi-phase field paradigm is used to capture interface evolution due to both short range corrosion reactions and mechanical crack growth, including their synergistic interaction. By simulating the evolving morphology of the electrolyte-metal interface, the underlying physical mechanisms can be resolved and the role of SCC defects (pits, cracks) can be investigated. The diffusion of metal ions away from the interface is also modelled to capture the interplay between diffusion- and activation-controlled corrosion. Moreover, we incorporate the role of passivation and the rupture of the passive film upon attaining a critical strain rate \citep{Parkins1987,JMPS2021}. The transport of hydrogen within the crystal lattice is predicted by means of an extended version of Fick's law and the impact of hydrogen content on the material toughness is accounted for by means of a quantum-mechanical approach \citep{Serebrinsky2004,CMAME2018}. Mechanical deformation is modelled using $J_2$ flow theory and we incorporate the role of mechanical fields in enhancing corrosion kinetics \citep{Gutman1998} and bulk hydrogen diffusion \citep{Sofronis1989,IJHE2016}. We numerically implement our generalised theory using the finite element method and demonstrate its predictive capabilities by analysing several representative case studies. First, two representative benchmarks are addressed to validate the pitting corrosion and hydrogen embrittlement predictive capabilities of the model. Then, numerical experiments are conducted to compare, qualitatively and quantitatively, model predictions against testing data and observations. A wide range of environments are considered, spanning the transition from hydrogen-driven to dissolution-assisted SCC. The competition and interplay between these two phenomena are investigated by: (i) simulating the failure of stainless steel samples under bending, (ii) predicting SCC thresholds in artificial and biologically-active seawater, and (iii) modelling the seminal experiments by \citet{Gruhl1984} on an AlZn5Mg3 alloy. A good agreement with experimental observations is attained.\\


\noindent \textit{Notation.}
We use lightface italic letters for scalars, e.g. $\phi$, upright bold letters for vectors, e.g. $\mathbf{u}$, and bold italic letters, such as $\bm{\sigma}$, for second and higher order tensors. First-, second-, and fourth-order tensors are in most cases respectively represented by small Latin, small Greek, and capital Latin letters. Inner products are denoted by a number of vertically stacked dots, corresponding to the number of indices over which summation takes place, such that $\bm{\sigma} : \bm{\varepsilon} = \sigma_{ij}\varepsilon_{ij}$, with indices referring to a Cartesian coordinate system. The gradient and the Laplacian are respectively denoted by $\nabla\mathbf{u}= u_{i,j}$ and $\nabla^2 \phi=\phi_{,ii}$. Finally, divergence is denoted by $\nabla\cdot\bm{\sigma}=\sigma_{ij,j}$, the trace of a second order tensor is written as $\text{tr} \,\bm{\varepsilon}=\varepsilon_{ii}$, and the deviatoric part of a tensor is given by $\bm{\sigma}'=\sigma_{ij}-\delta_{ij} \sigma_{kk}$, with $\delta_{ij}$ denoting the Kronecker delta. 

\section{Theory}
\label{Sec:Theory}

In this Section we formulate our generalised multi-phase-field theory, aimed at predicting the evolution of the aqueous electrolyte-solid metal interface due to corrosion, embrittlement and fracture processes. The theory refers to the electro-chemo-mechanical response of a domain $\Omega \subset {\rm I\!R}^n$ $(n \in[1,2,3])$ with external boundary $\partial \Omega\subset {\rm I\!R}^{n-1}$, on which the outwards unit normal is denoted as $\mathbf{n}$. For simplicity, isothermal conditions and isotropic material behaviour are assumed. We shall first define the kinematic variables (Section \ref{Sec:Kinematics}), proceed to derive the force balances using the principle of virtual work (Section \ref{Sec:PVW}), and then establish thermodynamically-consistent constitutive choices based on the free energy imbalance (Section \ref{Sec:imbalance}) and a generalised free energy definition (Section \ref{Sec:energy definition}). Our constitutive choices are detailed in Section \ref{Sec:constitutive theory} and a summary of the governing equations and of the interactions between the multiple elements of the model is given in Section \ref{Sec:SummaryTheory}.

\subsection{Kinematics and primal fields}
\label{Sec:Kinematics}

\begin{figure}[H]
\centering
\noindent\makebox[\textwidth]{%
\includegraphics[scale=0.18]{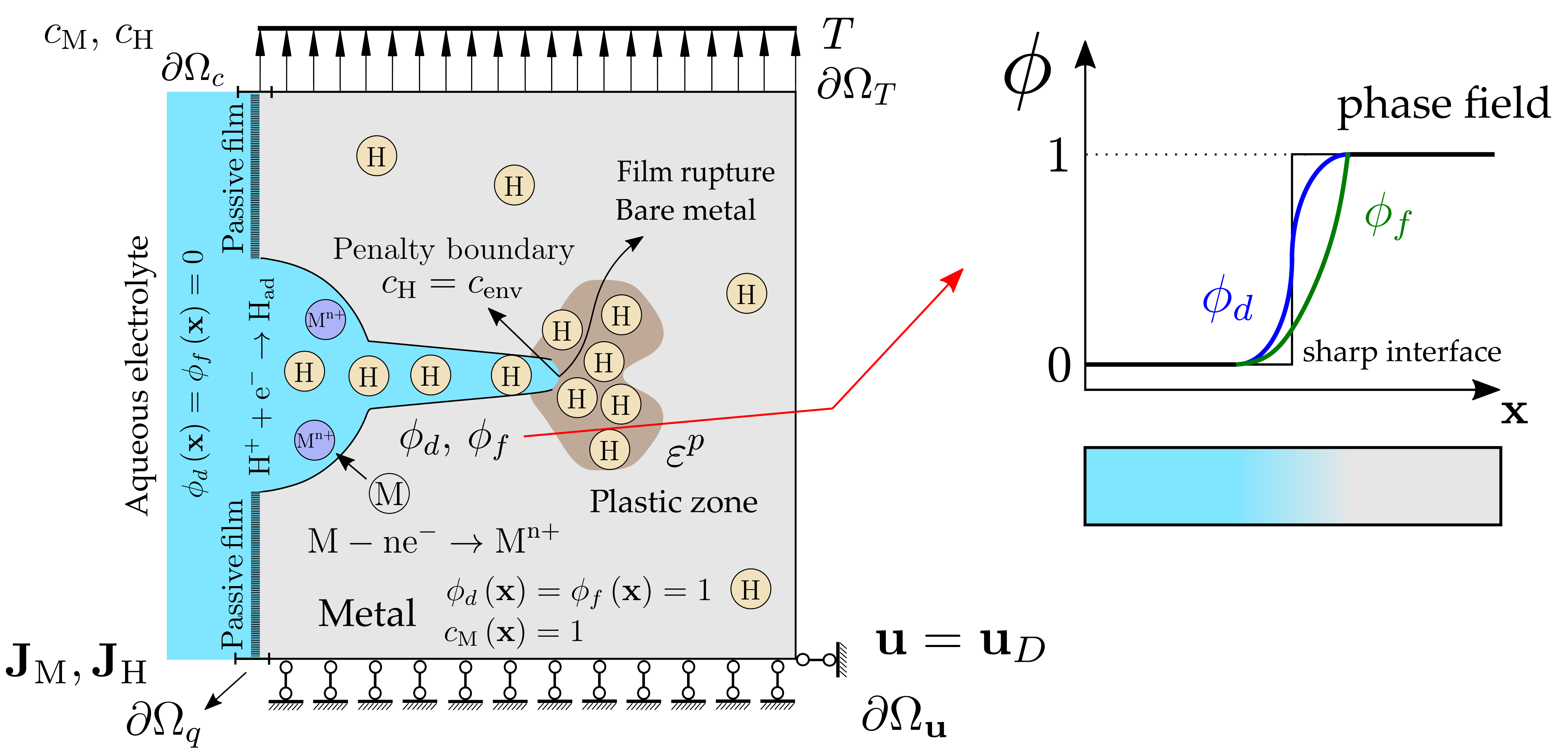}}
\caption{Schematic of the generalised SCC problem and the multi-phase-field framework proposed for modelling material dissolution (atom detachment), embrittlement and fracture. The model includes two separated phase field order parameters for the dissolution ($\phi_d$) and cracking ($\phi_f$) processes.}
\label{fig:SCCHE}
\end{figure}

As sketched in Fig. \ref{fig:SCCHE}, the primal fields are the displacement of the solid $\mathbf{u}$, the normalised concentration of metal ions in the electrolyte $c_\mathrm{M}$, the concentration of hydrogen atoms in the metal $c_\mathrm{H}$ and the two phase field variables: one describing the material dissolution process, $\phi_d$, and another one describing the rupture process, $\phi_f$.

Regarding the displacement field $\mathbf{u}$, the outer surface of the body can be decomposed into a part $\partial\Omega_u$, where the displacement is prescribed by Dirichlet-type boundary conditions, and a part $\partial\Omega_T$, where a traction $\mathbf{T}$ is prescribed by Neumann-type boundary conditions. Also, small strains are assumed, such that the total strain tensor $\bm{\varepsilon}$ reads
\begin{equation}
    \bm{\varepsilon} = \frac{1}{2}\left(\nabla\mathbf{u}^T+\nabla\mathbf{u}\right) \, .
\end{equation}

\noindent And the standard decomposition of strains into elastic and plastic components is adopted, such that:
\begin{equation}
    \bm{\varepsilon} = \bm{\varepsilon}^e+ \bm{\varepsilon}^p \, .
\end{equation}

Regarding the transport of ionic or atomic species, the external surface consists of two parts: $\partial \Omega_q$ where the metal ion flux $\mathbf{J}_{\mathrm{M}}$ or the hydrogen flux $\mathbf{J}_{\mathrm{H}}$ is known (Neumman-type boundary conditions), and $\partial \Omega_c$, where the metal ion concentration $c_{\mathrm{M}}$ or hydrogen concentration $c_\mathrm{H}$ is prescribed (Dirichlet-type boundary conditions). Note that we define $c_{\mathrm{M}}=c_{\mathrm{M}}'/c_{\mathrm{solid}}$ as the normalised concentration, ranging from 0 to 1, with $c_{\mathrm{M}}'$ being the molar concentration of the metal ion and $c_{\text{solid}}$ being the molar concentration of the metal ion in the solid. Besides, we adopt a penalty approach to enforce that the hydrogen concentration at the electrolyte-solid interface corresponds to that of the environment, $c_\text{H}=c_\text{env}$ \citep{CS2020}. This emulates how the newly created corrosion and crack surfaces are promptly exposed to the aqueous environment. The transport of species is driven by the chemical potential $\mu$ and, accordingly, we define a scalar field $\eta$ to determine the kinematics of composition changes \citep{Duda2018}, such that
\begin{equation}
    \dot{\eta} = \mu \, \, \, \, \, \, \text{and} \, \, \, \, \, \, \eta (\mathbf{\mathrm{x}}, t) = \int_0^t \mu (\mathbf{\mathrm{x}}, t) \, \text{d} t,  \, \, \, \text{ for both } \, \mu_\mathrm{M} \, \text{ and }  \, \mu_\mathrm{H} \, .
\end{equation}

Finally, the morphology of the evolving electrolyte-metal interface is characterised by two phase field variables; a dissolution phase field $\phi_d$ that captures the short-range corrosion reactions (atoms detaching), and a fracture phase field $\phi_f$, which captures the creation of new surfaces due to the breaking of atomic bonds. In both cases, see Fig. \ref{fig:SCCHE}, we assume that $\phi_d=\phi_f=1$ corresponds to the solid, undamaged material points, while $\phi_d=\phi_f=0$ in the electrolyte. As described below, both formulations are non-local, involving their respective phase field gradients ($\nabla \phi_d$, $\nabla \phi_f$) and micro-stress quantities work conjugate to those gradients ($\mathbf{\zeta}_d$, $\mathbf{\zeta}_f$). The formulation allows for defining phase field micro-tractions, $f_d$ and $f_f$, and also for the prescription of initial conditions and Dirichlet-type boundary conditions on the phase field variables. Thus, the interface morphology at time $t=0$ can be set by defining a suitable geometry or by defining an initial condition on the phase field variables.

\subsection{Principle of virtual work and force balances}
\label{Sec:PVW}

Let us now derive the local force balances from the Principle of Virtual Work (PVW). In consistency with the kinematic fields described in Section \ref{Sec:Kinematics}, we define a set of virtual fields as $\left( \delta \mathbf{u}, \, \delta \phi_d, \, \delta \eta_\mathrm{M},\, \delta \phi_f,  \, \delta \eta_\mathrm{H} \right)$. For an arbitrary part $\Omega$, considering the boundary definitions in Fig. \ref{fig:SCCHE}, the internal virtual work $\delta \mathcal{W}^{\text{int}}$ and external virtual work $\delta \mathcal{W}^{\text{ext}}$ are given by
\begin{equation}\label{eq:intwork}
    \begin{aligned}
    \delta \mathcal{W}^{\text{int}}&\left(\Omega;\delta \mathbf{u}, \, \delta \phi_d, \, \delta \eta_\mathrm{M},\, \delta \phi_f,  \, \delta \eta_\mathrm{H}\right)=\int_\Omega \left(
    \bm{\sigma} : \mathrm{sym} \nabla \delta \mathbf{u} + \omega_d \, \delta \phi_d + \mathbf{\zeta}_d \cdot \nabla \delta \phi_d \right.\\
    \phantom{=\;\;}
    & \left. - \mathbf{J}_\mathrm{M} \cdot \nabla \delta \eta_\mathrm{M} + \, \omega_f \,\delta \phi_f + \mathbf{\zeta}_f \cdot \nabla \delta \phi_f - \mathbf{J}_\mathrm{H} \cdot \nabla \delta \eta_\mathrm{H} \right) \, \text{d} V
    \end{aligned}
\end{equation}
and 
\begin{equation}\label{eq:extwork}
    \begin{aligned}
    \delta \mathcal{W}&^{\text{ext}}\left(\Omega;\delta \mathbf{u}, \, \delta \phi_d, \, \delta \eta_\mathrm{M},\, \delta \phi_f,  \, \delta \eta_\mathrm{H}\right)=\int_{\partial\Omega} \left( \mathbf{T} \cdot \delta \mathbf{u} + f_d\, \delta \phi_d + f_f \delta \phi_f + q_\mathrm{M} \, \delta \eta_\mathrm{M}\right.\\
    \phantom{=\;\;}
    & \left. + q_\mathrm{H} \, \delta \eta_\mathrm{H} \right) \, \text{d} S + \int_\Omega \left( \mathbf{b}^{\text{ext}} \cdot \delta \mathbf{u} + f_d^{\text{ext}}\, \delta \phi_d + f_f^{\text{ext}} \delta \phi_f + q_\mathrm{M}^{\text{ext}} \, \delta \eta_\mathrm{M} + q_\mathrm{H}^{\text{ext}} \, \delta \eta_\mathrm{H}\right)  \, \text{d} V
    \end{aligned}
\end{equation}

\noindent where $\bm{\sigma}$ is the Cauchy stress tensor, $\mathbf{b}$ is the body force vector and $\omega_d$ and $\omega_f$ denote the microstress quantities work conjugate to $\phi_d$ and $\phi_f$, respectively. Note that, following \citet{Duda2018}, we decompose the external work quantities into the contact interactions on $\partial\Omega$ and the distant interactions, which account for the interactions with all bodies external to $\Omega$. The superscript $\Hsquare^{\text{ext}}$ denotes the variables in the latter group. The interactions at a distance can be further divided into a non-inertial part that accounts for interactions of bodies accessible to observation, and an inertial part account for interactions with the remaining bodies. Here, we assume that $\mathbf{b}^{\text{ext}}$, $f_d^{\text{ext}}$ and $f_f^{\text{ext}}$ are purely non-inertial but we decompose $q_\mathrm{M}$ and $q_\mathrm{H}$ as follows:
\begin{equation}\label{eq:qMqH}
   q_\mathrm{M}^{\text{ext}}=q_\mathrm{M}^{\text{ni}}-\dot{c}_\mathrm{M}c_\mathrm{solid},\,\,\,\,\,\,\,\,\,q_\mathrm{H}^{\text{ext}}=q_\mathrm{H}^{\text{ni}}-\dot{c}_\mathrm{H}
\end{equation}
\noindent where the superscript $\Hsquare^{\text{ni}}$ is used to denote the non-inertial parts. The second term on the right hand sides of (\ref{eq:qMqH}) are the inertial parts of $q_\mathrm{M}$ and $q_\mathrm{H}$, and correspond to the solute number densities. By inserting (\ref{eq:qMqH}) into (\ref{eq:extwork}), the external virtual work $\delta \mathcal{W}^{\text{ext}}$ is naturally decomposed into a non-inertial part $\delta \mathcal{W}^{\text{ni}}$ and an inertial part $\delta \mathcal{W}^{\text{in}}$, as follows
\begin{equation}\label{eq:extwork2}
    \begin{aligned}
    \delta \mathcal{W}&^{\text{ext}}=\delta \mathcal{W}^{\text{ni}} +  \delta \mathcal{W}^{\text{in}}
    \end{aligned}
\end{equation}
with
\begin{equation}\label{eq:extworkni}
    \begin{aligned}
    \delta \mathcal{W}&^{\text{ni}}\left(\Omega;\delta \mathbf{u}, \, \delta \phi_d, \, \delta \eta_\mathrm{M},\, \delta \phi_f,  \, \delta \eta_\mathrm{H}\right)=\int_{\partial\Omega} \left( \mathbf{T} \cdot \delta \mathbf{u} + f_d\, \delta \phi_d + f_f \delta \phi_f + q_\mathrm{M} \, \delta \eta_\mathrm{M}\right.\\
    \phantom{=\;\;}
    & \left. + q_\mathrm{H} \, \delta \eta_\mathrm{H} \right) \, \text{d} S + \int_\Omega \left( \mathbf{b}^{\text{ni}} \cdot \delta \mathbf{u} + f_d^{\text{ni}}\, \delta \phi_d + f_f^{\text{ni}} \delta \phi_f + q_\mathrm{M}^{\text{ni}} \, \delta \eta_\mathrm{M} + q_\mathrm{H}^{\text{ni}} \, \delta \eta_\mathrm{H}\right)  \, \text{d} V
    \end{aligned}
\end{equation}
and 
\begin{equation}\label{eq:extworkin}
    \begin{aligned}
    \delta \mathcal{W}&^{\text{in}}\left(\Omega;\delta \mathbf{u}, \, \delta \phi_d, \, \delta \eta_\mathrm{M},\, \delta \phi_f,  \, \delta \eta_\mathrm{H}\right)= -\int_\Omega \left(\dot{c}_\mathrm{M}c_\mathrm{solid} \delta \eta_\mathrm{M} + \dot{c}_\mathrm{H} \delta \eta_\mathrm{H}\right) \, \text{d} V
    \end{aligned}
\end{equation}

We proceed to introduce the principle of virtual work, such that
\begin{equation}\label{eq:virtual work}
    \begin{aligned}
   \delta \mathcal{W}^{\text{int}}\left(\Omega;\delta \mathbf{u}, \, \delta \phi_d, \, \delta \eta_\mathrm{M},\, \delta \phi_f,  \, \delta \eta_\mathrm{H}\right)=\delta \mathcal{W}^{\text{ext}}\left(\Omega;\delta \mathbf{u}, \, \delta \phi_d, \, \delta \eta_\mathrm{M},\, \delta \phi_f,  \, \delta \eta_\mathrm{H}\right)
    \end{aligned}
\end{equation}

Now, let us neglect all non-inertial parts such that $\mathbf{b}^{\text{ni}}=\mathbf{0}$ and $f_d^{\text{ni}}=f_f^{\text{ni}}=q_\mathrm{M}^{\text{ni}}=q_\mathrm{H}^{\text{ni}}=0$. Then, combining (\ref{eq:intwork})-(\ref{eq:virtual work}) and following Gauss' divergence theorem, the equilibrium equations on $\Omega$ are obtained as,
\begin{equation}\label{eq:PVWb}
    \begin{aligned}
   & \int_\Omega \Bigl\{ \left( \nabla \cdot \bm{\sigma} \right) \cdot \delta \mathbf{u} +  \left( \nabla \cdot \mathbf{\zeta}_d - \omega_d \right) \delta \phi_d -  \left( \nabla \cdot \mathbf{J}_\mathrm{M}  + \dot{c}_\mathrm{M}c_\mathrm{solid} \right)  \delta \eta_\mathrm{M}\Bigr.\\
    \phantom{=\;\;}
    & \,\,\,\,\,\,\,\, \Bigl. +  \left( \nabla \cdot \mathbf{\zeta}_f - \omega_f \right) \delta \phi_f - \left( \nabla \cdot \mathbf{J}_\mathrm{H}  + \dot{c}_\mathrm{H} \right)  \delta \eta_\mathrm{H} \Bigr\} \, \text{d} V \\
    & = \int_{\partial \Omega} \Bigl\{ \left( \bm{\sigma} \cdot \mathbf{n}- \mathbf{T} \right) \cdot \delta \mathbf{u} + \left( \mathbf{\zeta}_d \cdot \mathbf{n}  - f_d \right) \delta \phi_d + \left( \mathbf{J}_\mathrm{M} \cdot \mathbf{n}  + q_\mathrm{M} \right) \delta \eta_\mathrm{M}  \Bigr.\\
    & \,\,\,\,\,\,\,\, \Bigl. + \left( \mathbf{\zeta}_f \cdot \mathbf{n}  - f_f \right) \delta \phi_f + \left( \mathbf{J}_\mathrm{H} \cdot \mathbf{n}  + q_\mathrm{H} \right) \delta \eta_\mathrm{H}  \Bigr\} \, \text{d} S
    \end{aligned}
\end{equation}

Since the left-hand side of (\ref{eq:PVWb}) must vanish for arbitrary variations, the equilibrium equations in $\Omega$ are obtained as,
\begin{equation}\label{eq:LocalBalance}
    \begin{aligned}
    \nabla \cdot \bm{\sigma} = \mathbf{0} \\
    \nabla \cdot \mathbf{\zeta}_d - \omega_d = 0 \\
    \dot{c}_\mathrm{M}c_\mathrm{solid} + \nabla \cdot \mathbf{J}_\mathrm{M} = 0\\
    \nabla \cdot \mathbf{\zeta}_f - \omega_f = 0 \\
    \dot{c}_\mathrm{H} + \nabla \cdot \mathbf{J}_\mathrm{H} = 0
    \end{aligned}
\end{equation}

\noindent and the right-hand side of (\ref{eq:PVWb}) provides the corresponding set of boundary conditions on $\partial \Omega$,
\begin{equation}\label{eq:LocalBalance_BC}
    \begin{aligned}
    \mathbf{T} = \bm{\sigma} \cdot \mathbf{n} \\
    f_d = \mathbf{\zeta}_d \cdot \mathbf{n} \\
    q_\mathrm{M} = -\mathbf{J}_\mathrm{M} \cdot \mathbf{n}\\
    f_f = \mathbf{\zeta}_f \cdot \mathbf{n} \\
    q_\mathrm{H} = -\mathbf{J}_\mathrm{H} \cdot \mathbf{n}\\
    \end{aligned}
\end{equation}

\subsection{Energy imbalance}
\label{Sec:imbalance}

The first two laws of thermodynamics for a continuum body within a dynamical process of specific internal energy $\mathscr{E}$ and specific entropy $\mit \Lambda$ read \citep{Gurtin2010},
\begin{equation}\label{eq:thermodynamics laws}
    \begin{aligned}
    & \frac{d}{dt} \int_\Omega \mathscr{E} \, \text{d} V = \dot{\mathcal{W}}^{\text{ni}} - \int_{\partial \Omega} \mathbf{Q} \cdot \mathbf{n} \, \text{d} S + \int_\Omega Q \, \text{d} V \\
    & \frac{d}{dt} \int_\Omega {\mit \Lambda} \, \text{d} V \geqslant - \int_{\partial \mathrm{\Omega}} \frac{ \mathbf{Q} }{T} \cdot \mathbf{n} \, \text{d} S +  \int_\mathrm{\Omega} \frac{Q}{T} \, \text{d} V
    \end{aligned}
\end{equation}

\noindent where $\dot{\mathcal{W}}^{\text{ni}}$ is the non-inertial part of external power, $\mathbf{Q}$ is the heat ﬂux and $Q$ is the heat absorption. Here we limit our discussion within isothermal condition $T=T_0$ such that (\ref{eq:thermodynamics laws}) can be re-written as:
\begin{equation}\label{eq:thermodynamics laws2}
    \begin{aligned}
    \frac{d}{dt} \int_\Omega \psi \, \text{d} V \leqslant \dot{\mathcal{W}}^{\text{ni}}
    \end{aligned}
\end{equation}
\noindent where $\psi=\mathscr{E}-T{\mit \Lambda}$ is the free energy density. Replacing the virtual fields $\left( \delta \mathbf{u}, \, \delta \phi_d, \, \delta \eta_\mathrm{M}, \, \delta \phi_f, \, \delta \eta_\mathrm{H} \right)$ by realisable velocity fields $( \dot{\mathbf{u}}, \, \dot{\phi}_d, \, \mu_\mathrm{M}, \, \dot{\phi}_f, \, \mu_\mathrm{H} )$ in (\ref{eq:intwork}) and (\ref{eq:extwork}), one can reach
\begin{equation}\label{eq:Imbalance}
    \frac{d}{dt} \int_\Omega \psi \, \text{d} V \leqslant \int_{\partial \Omega} \left( \mathbf{T} \cdot \dot{\mathbf{u}} + f_d\, \dot{\phi}_d + f_f \, \dot{\phi}_f + q_\mathrm{M} \, \mu_\mathrm{M} + q_\mathrm{H} \, \mu_\mathrm{H} \right) \, \text{d} S
\end{equation}

Employing the divergence theorem and recalling the local balance equations (\ref{eq:LocalBalance})-(\ref{eq:LocalBalance_BC}), we find the equivalent point-wise version of (\ref{eq:Imbalance}) as,
\begin{equation}\label{eq:Imbalance2}
    \begin{aligned}
    \dot{\psi} & - \bm{\sigma} :  \dot{\bm{\varepsilon}} - \left( \omega_d \dot{\phi}_d + \mathbf{\zeta}_d \cdot \nabla \dot{\phi}_d \right) + \left( \frac{}{}\mu_\mathrm{M} \dot{c}_\mathrm{M} \, c_\mathrm{solid} - \mathbf{J}_\mathrm{M} \cdot \nabla \mu_\mathrm{M} \right) \\
    & - \left( \omega_f \dot{\phi}_f + \mathbf{\zeta}_f \cdot \nabla \dot{\phi}_f \right) + 
    \left( \frac{}{}\mu_\mathrm{H} \dot{c}_\mathrm{H} - \mathbf{J}_\mathrm{H} \cdot \nabla \mu_\mathrm{H} \right)
    \leqslant 0
    \end{aligned}
\end{equation}

\subsection{Free energy density}
\label{Sec:energy definition}

We proceed to define the functional form of free energy density $\psi$ in (\ref{eq:Imbalance2}). The free energy can be decomposed into the mechanical free energy density $\psi^m$, the chemical free energy density $\psi^{ch}$, the dissolution interface energy density $\psi^i$, the fracture interface energy density $\psi^f$, and the hydrogen-chemical energy density $\psi^H$, such that
\begin{equation}
    \begin{aligned}
    \psi = \psi^m + \psi^{ch} + \psi^i + \psi^f + \psi^H 
    \end{aligned}
\end{equation}

\subsubsection{Mechanical free energy density}

As shown in Fig. \ref{fig:SCCHE}, in combining phase field corrosion and phase field fracture concepts, we choose to retain their original form and thus define different degradation functions for dissolution ($h_d$) and fracture ($h_f$), such that
\begin{equation}
    \begin{aligned}
    h_d \left(\phi_d \right)= -2 \phi_d^3 + 3 \phi_d^2  \,\,\,\,\,\,\,\,\,\, \text{and} \,\,\,\,\,\,\,\,\,\,  h_f \left(\phi_f \right)= \phi_f^2 \, 
    \end{aligned}
\end{equation}

Subsequently, $\psi^m$ is defined as:
\begin{equation}
    \begin{aligned}
    \psi^m = h_d\left(\phi_d \right) h_f\left(\phi_f \right) \psi_0^m - K \overline{V}_\mathrm{H} \left( c_\mathrm{H} - c_{\mathrm{H}0} \right) \text{tr} \, \bm{\varepsilon} 
    \end{aligned}
\end{equation}
\noindent where the first term on the right hand side is aimed at capturing the joint effect of dissolution and damage on the mechanical free energy density. Here, $K$ is the bulk modulus, $\overline{V}_\mathrm{H}$ is the partial molar volume of hydrogen in solid solution, and $c_{\mathrm{H}0}$ is the reference hydrogen concentration. For an elastic-plastic solid, the mechanical free energy density $\psi_0^m$ of the undamaged solid is given by:
\begin{equation}
    \begin{aligned}
    \psi_0^m = \psi_0^{e} + \psi_0^{p} = \frac{1}{2} \bm{\varepsilon}^e : \bm{C}^e : \bm{\varepsilon}^e + \int_0^t \bm{\sigma}_0 : \dot{\bm{\varepsilon}}^p \, \text{d}t
    \end{aligned}
\end{equation}
\noindent where $\psi_0^{e}$ and $\psi_0^{p}$ respectively denote the elastic and plastic parts of the (undamaged) mechanical free energy density, $\bm{C}^e$ is the linear elastic stiffness matrix, and $\bm{\sigma}_0$ is the Cauchy stress tensor for the undamaged solid.

\subsubsection{Electrochemical energy density}

We follow the KKS model \citep{Kim1999} to define the chemical free energy density $\psi^{ch}$, assuming that each material point is a mix of both solid and liquid phases with different concentrations but similar chemical potentials, such that 
\begin{equation}\label{eq:Phi_E1}
    \psi^{ch} = h_d \left( \phi_d \right) \psi^{ch,S} + \left[ 1-  h_d\left(\phi_d\right) \right] \psi^{ch,L} + w g \left( \phi_d \right)
\end{equation}
and 
\begin{equation}\label{eq:c}
    c_\mathrm{M} = h_d \left( \phi_d \right) c_\mathrm{M}^S + \left[ 1 - h_d \left( \phi_d \right) \right] c_\mathrm{M}^L
\end{equation}

\begin{equation}\label{eq:dc}
    \frac{\partial \psi^{ch,S} \left(  c_\mathrm{M}^S\right)}{\partial  c_\mathrm{M}^S} =  \frac{\partial \psi^{ch,L} \left(  c_\mathrm{M}^L\right)}{\partial  c_\mathrm{M}^L}
\end{equation}
\noindent where $\psi^{ch,S}$ and $\psi^{ch,L}$ denote the free energy density terms associated with the solid and liquid phases, $c_\mathrm{M}^S$ and $c_\mathrm{M}^L$ denote the normalized concentrations of the co-existing solid and liquid phases, and $w$ is the height of the double well potential $g \left( \phi_d \right)={\phi_d}^2 (1 - \phi_d)^2$. $\psi^{ch,S}$ and $\psi^{ch,L}$ are defined as \citep{Mai2016}:
\begin{equation}\label{eq:fsfl}
    \psi^{ch,S} = A ( c_\mathrm{M}^S - c_\mathrm{Se})^2 \,\,\,\,\,\,\,\, \text{and} \,\,\,\,\,\,\,\, \psi^{ch,L} = A ( c_\mathrm{M}^L - c_\mathrm{Le})^2
\end{equation}

\noindent where $A$ is the free energy density curvature, and $c_\mathrm{Se}=c_\mathrm{solid}/c_\mathrm{solid}=1$ and $c_\mathrm{Le}=c_\mathrm{sat}/c_\mathrm{solid}$ respectively denote the normalised equilibrium concentrations for the solid and liquid phases. Combining (\ref{eq:Phi_E1})-(\ref{eq:fsfl}) renders,
\begin{equation}\label{eq:Phi_E}
    \psi^{ch} = A \left[ c_\mathrm{M} - h_d\left(\phi_d\right)(c_\mathrm{Se} - c_\mathrm{Le}) - c_\mathrm{Le} \right]^2 + w \,{\phi_d}^2 \left( 1 - \phi_d \right)^2 \, .
\end{equation}

\subsubsection{Dissolution interface energy density}

The dissolution interface energy density $\psi^i$ is defined as a function of the gradient of the corrosion phase field order parameter $\phi_d$ as:
\begin{equation}\label{eq:phi_D}
    \psi^i = \frac{\alpha}{2} |\nabla \phi_d|^2 
\end{equation}
\noindent where $\alpha$ is the gradient energy coefficient. The parameters $\alpha$ and $w$ in the electrochemical system (\ref{eq:Phi_E})-(\ref{eq:phi_D}) are related to the interface energy $\gamma$ and its thickness $\ell_d$, rendering
\begin{equation}\label{eq:gammal}
    \gamma = \sqrt{\frac{\alpha w}{18}} \,\,\,\,\,\,\,\, \text{and} \,\,\,\,\,\,\,\, \ell_d = a^* \sqrt{\frac{2 \alpha }{w}} 
\end{equation}
\noindent where $a^*=2.94$ is a constant parameter corresponding to the definition of the interface region $0.05 < \phi < 0.95$ \citep{Abubakar2015}.

\subsubsection{Fracture interface energy density}
\label{Sec:FractureInterfaceEnergyDensity}

The phase field description of the fracture problem is aimed at regularising Griffith's energy balance and determine the evolution of $\phi_f$ based on the thermodynamics of fracture \citep{Bourdin2008}. Accordingly, for a material toughness $G_c$, the fracture interface energy density $\psi^f$ is given by,
\begin{equation}\label{eq:phi_F}
    \psi^f = G_c \left(c_\mathrm{H}\right) \left[ \frac{\ell_f}{2} \, |\nabla \phi_f|^2 + \frac{1}{2\,\ell_f} \left(1-\phi_f\right)^2 \right]
\end{equation}
\noindent where $\ell_f$ is the length scale parameter, governing the size of the regularised region. As shown in (\ref{eq:phi_F}), the role of hydrogen in degrading the fracture resistance of the solid is introduced by making the critical energy release rate sensitive to the hydrogen content. The formulation is general, in that it can incorporate any definition of the toughness degradation to accommodate any mechanistic interpretation. One possibility, as presented by \citet{CMAME2018}, is to adopt an atomistically-informed approach, by which the fracture energy degradation is related to atomistic computations of surface energy sensitivity to hydrogen coverage, $\theta$. The hydrogen coverage at an interface can be estimated by means of the Langmuir–McLean isotherm \citep{Serebrinsky2004}:
\begin{equation}\label{eq:theta}
    \theta =\frac{c_\mathrm{H}}{c_\mathrm{H} + \text{exp}\left(\frac{- \Delta g_b^0}{RT}\right)}
\end{equation}

\noindent where $R$ is the universal gas constant and $\Delta g_b^0$ is the interface binding energy. For example, for hydrogen assisted failures driven by grain boundary decohesion \citep{Harris2018}, $\Delta g_b^0$ corresponds to the binding energy of grain boundaries, which is on the order of 30 kJ/mol. As noted by \citet{CMAME2018}, atomistic calculations of surface energy degradation with increasing hydrogen coverage can be approximated with a linear trend, such that:
\begin{equation}\label{eq:G_c}
     G_c \left(c_\mathrm{H}\right) =G_c \left(\theta\right)= \left(1-\chi \theta \right) G_c \left(0\right)
\end{equation}
\noindent where $\chi$ is the so-called hydrogen damage coefficient. For Fe-based materials, $\chi=0.89$ provides the best fit to the data by \citet{Jiang2004a}, while $\chi=0.41$ provides the best fit to the DFT simulations on Ni by \citet{Alvaro2015}. In this work, we will make use of this atomistically-informed approach to provide fully predictive estimates (i.e., without the need for experimental calibration) but we will also consider phenomenological approaches and discuss their differences.

\subsubsection{Hydrogen-chemical energy density}

Hydrogen transport is driven by chemical potential gradients. For a given number of hydrogen sites $N_\mathrm{H}$ with occupancy $\theta_\mathrm{H}=c_\mathrm{H}/N_\mathrm{H}$, the hydrogen-chemical energy density is given by \citep{JMPS2020}:
\begin{equation}\label{eq:phi_H}
    \psi^H = \mu_0 c_{\mathrm{H}} + RTN_\mathrm{H}\left[\theta_\mathrm{H}\ln{\theta_\mathrm{H}}+(1-\theta_\mathrm{H})\ln{(1-\theta_\mathrm{H})}\right]
\end{equation}
\noindent where $\mu_0$ is the reference chemical potential. The formulation is general in that it can refer to the transport of lattice hydrogen or to the transport of diffusible hydrogen, upon appropriate choices of the hydrogen diffusion coefficient $D_\mathrm{H}$. Thus, the role of microstructural trapping sites can be accounted for in the present formulation, both in terms of their involvement in the fracture process (see Section \ref{Sec:FractureInterfaceEnergyDensity}) and in their role in reducing diffusion rates, through a suitable choice of $D_\mathrm{H}$. For simplicity, we do not capture the effect of an evolving trap density, as it is the case of dislocation trap sites, but such extension is straightforward (see \citealp{IJP2021}).

\subsection{Constitutive theory}
\label{Sec:constitutive theory}

Let us now outline our constitutive choices, consistent with the free energy imbalance (Section \ref{Sec:imbalance}) and the free energy definition (Section \ref{Sec:energy definition}). First, following Anand and co-workers (see, e.g. \citealp{Narayan2019,Anand2019}), the dissipative nature of the corrosion and fracture processes is emphasised by decomposing the dissolution and fracture micro-stresses into energetic and dissipative parts:
\begin{equation}\label{eq:omegadf}
    \omega_d = \omega_d^{\text{en}} + \omega_d^{\text{dis}}\,\,\,\,\,\,\,\, \text{and} \,\,\,\,\,\,\,\, \omega_f = \omega_f^{\text{en}} + \omega_f^{\text{dis}} \, .
\end{equation}

Accordingly, the constitutive relations can be derived by fulfilling the free energy
imbalance:
\begin{equation}\label{eq:constitutive}
    \begin{aligned}
    &\left(\frac{\partial \,\psi}{\partial \bm{\varepsilon}} - \bm{\sigma} \right): \dot{\bm{\varepsilon}}  + \left(\frac{\partial \,\psi}{\partial \phi_d}-\omega_d^{\text{en}} \right) \dot{\phi}_d + \left(\frac{\partial\, \psi}{\partial \nabla \phi_d}-\mathbf{\zeta}_d \right) \nabla \dot{\phi}_d\\ 
    & \,\,\,\,\,\, + \left[ \left(\frac{\partial \,\psi}{\partial {c}_\mathrm{M}}+\mu_\mathrm{M}c_\mathrm{solid} \right)\dot{c}_\mathrm{M} - \mathbf{J}_\mathrm{M} \cdot \nabla \mu_\mathrm{M} \right] + \left(\frac{\partial \,\psi}{\partial \phi_f}-\omega_f^{\text{en}} \right) \dot{\phi}_f \\
    & \,\,\,\,\,\, + \left(\frac{\partial \,\psi}{\partial \nabla \phi_f}-\mathbf{\zeta}_f \right) \nabla \dot{\phi}_f + \left[ \left(\frac{\partial\, \psi}{\partial {c}_\mathrm{H}}+\mu_\mathrm{H} \right)\dot{c}_\mathrm{H} - \mathbf{J}_\mathrm{H} \cdot \nabla \mu_\mathrm{H} \right] \leqslant 0
    \end{aligned}
\end{equation}
\noindent with free energy $\psi$ being defined, in agreement with Section \ref{Sec:energy definition}, as follows:
\begin{equation}\label{eq:Free_energy}
  \begin{aligned}
  \psi =& \psi\left(\bm{\varepsilon}, \phi_d, c_\mathrm{M}, \phi_f, c_\mathrm{H} \right) = \underbrace{ \left[h_d\left(\phi_d \right) h_f\left(\phi_f \right) \psi_0^m- K \overline{V}_\mathrm{H} \left( c_\mathrm{H} - c_{\mathrm{H}0} \right) \text{tr} \, \bm{\varepsilon}\right]}_{\psi^m} \\
  & + \underbrace{A \left[ c_\mathrm{M} - h_d\left(\phi_d\right)(c_\mathrm{Se} - c_\mathrm{Le}) - c_\mathrm{Le} \right]^2 + w \,{\phi_d}^2 \left( 1 - \phi_d \right)^2}_{\psi^{ch}} \\
  & + \underbrace{\frac{\alpha}{2} |\nabla \phi_d|^2}_{\psi^i} + \underbrace{ G_c \left(c_\mathrm{H}\right) \left[ \frac{\ell_f}{2} \, |\nabla \phi_f|^2 + \frac{1}{2\,\ell_f} \left(1-\phi_f\right)^2 \right]}_{\psi^f}\\
  & + \underbrace{\mu_0 c_{\mathrm{H}} + RTN_\mathrm{H}\left[\theta_\mathrm{H}\ln{\theta_\mathrm{H}}+(1-\theta_\mathrm{H})\ln{(1-\theta_\mathrm{H})}\right]}_{\psi^H}
  \end{aligned}
\end{equation}

\noindent Here, one should note that $\omega_d^{\text{dis}}$ and $\omega_f^{\text{dis}}$ must obey the following residual dissipation inequality,
\begin{equation}\label{eq:imblance_dis} 
  \begin{aligned}
  \omega_d^{\text{dis}} \geqslant 0 \,\,\,\,\,\,\,\, \text{and} \,\,\,\,\,\,\,\,\omega_f^{\text{dis}} \geqslant 0
  \end{aligned}
\end{equation}

The general formulations of the dissipative microstress quantities are given by:
\begin{equation}\label{eq:Omega_ddis}
    \omega_d^{\text{dis}} = \hat{a}\left(\bm{\varepsilon}, \, \phi_d, \, \nabla \phi_d, \, \eta_\mathrm{M},\, \phi_f,  \, \nabla \phi_f, \, \eta_\mathrm{H}\right) + \hat{b}\left(\bm{\varepsilon}, \, \phi_d, \, \nabla \phi_d, \, \eta_\mathrm{M},\, \phi_f,  \, \nabla \phi_f, \, \eta_\mathrm{H}\right) \dot{\phi}_d.
\end{equation}
\begin{equation}\label{eq:Omega_fdis}
    \omega_f^{\text{dis}} = \hat{\alpha}\left(\bm{\varepsilon}, \, \phi_d, \, \nabla \phi_d, \, \eta_\mathrm{M},\, \phi_f,  \, \nabla \phi_f, \, \eta_\mathrm{H}\right) + \hat{\beta}\left(\bm{\varepsilon}, \, \phi_d, \, \nabla \phi_d, \, \eta_\mathrm{M},\, \phi_f,  \, \nabla \phi_f, \, \eta_\mathrm{H}\right) \dot{\phi}_f.
\end{equation}
\noindent where $\hat{a}$, $\hat{b}$, $\hat{\alpha}$, and $\hat{\beta}$ are non-negative response functions to ensure satisfying of (\ref{eq:imblance_dis}). Particularly, the conditions $\hat{b}=0$ and $\hat{\beta}=0$ correspond to rate-independent processes. For simplicity, we follow \cite{Duda2015} and assume $\hat{a}=\hat{\alpha}=0$.

\subsubsection{Chemo-elastoplasticity}

The constitutive prescriptions for the mechanical behaviour of the solid are derived as follows. 
According to (\ref{eq:constitutive}) and (\ref{eq:Free_energy}), the Cauchy stress tensor can be derived as:
\begin{equation}\label{eq:CauchyStress}
    \bm{\sigma} = \frac{\partial \, \psi}{\partial \bm{\varepsilon}} = h_d\left(\phi_d \right) h_f\left(\phi_f \right)\bm{C^{\mathrm{ep}}}:\bm{\varepsilon} - K \overline{V}_\mathrm{H} \left( c_\mathrm{H} - c_{\mathrm{H}0} \right) \text{tr} \, \bm{I}
\end{equation}
\noindent where $\bm{C^{\mathrm{ep}}}=\partial \bm{\sigma} / \partial \bm{\varepsilon}$ is the consistent material Jacobian. The second term, corresponding to lattice dilation, is considered to have a negligible effect in hydrogen embrittlement phenomena \citep{Hirth1980} and is subsequently omitted.

The work hardening behaviour of the solid is assumed to be characterised by an isotropic power law, such that the relation between the flow stress $\sigma$ and the equivalent plastic strain $\varepsilon^p$ is given by,
\begin{equation}\label{eq:plastic}
    \sigma  = \sigma_y \left(1 + \frac{E \varepsilon^p}{\sigma_y} \right)^N
\end{equation}
\noindent where $E$ is Young’s modulus, $\sigma_y$ is the initial yield stress and $N$ is the strain hardening exponent ($0 \leqslant N \leqslant 1$).

For simplicity, we have chosen to describe the constitutive behaviour of the solid using $J_2$ plasticity theory. The framework can readily be incorporated to capture the role of crystal anisotropy \citep{Castelluccio2018,Kumar2020,Tondro2021} and dislocation hardening due to plastic strain gradients \citep{Komaragiri2008,EJMAS2019,JMPS2019}. The former is mostly relevant for predicting pit nucleation and the growth of very small cracks (smaller than a characteristic microstructural length). While strain gradient effects become most important in pre-cracked samples exposed to anodic environments, as strain gradient hardening will lead to higher hydrogen contents.

\subsubsection{A phase field description of metallic corrosion}
\label{subsec:PhaseFieldCorr}

The evolution of the corrosion front is characterised by the dissolution phase field microstress vector $\mathbf{\zeta}_d$, work conjugate to the phase field gradient,
\begin{equation}\label{eq:zeta}
    \mathbf{\zeta}_d = \frac{\partial \,\psi}{\partial \nabla \phi_d} = \alpha \nabla^2 \phi_d \, ,
\end{equation}

\noindent and the scalar micro-stress $\omega_d$, work conjugate to the dissolution phase field $\phi_d$. In agreement with (\ref{eq:Free_energy}), the energetic part of $\omega_d$ is defined as follows:
\begin{equation}\label{eq:omega}
    \begin{aligned}
    \omega_d^{\text{en}} &= \frac{\partial \,\psi}{\partial \phi_d} 
    = - 2 A \left[c_\mathrm{M} - h_d\left(\phi_d\right)(c_\mathrm{Se} - c_\mathrm{Le}) - c_\mathrm{Le} \right] (c_\mathrm{Se} - c_\mathrm{Le}) h_d'\left(\phi_d \right) \\
    & + w \left(4 \phi_d^3 + 2 \phi_d - 6 \phi_d^2\right) + h_d'\left(\phi_d \right) h_f\left(\phi_f \right) \psi_0^m
    \end{aligned}
\end{equation}

The last term in (\ref{eq:omega}) makes corrosion kinetics sensitive to the mechanical fields, as shown by \citet{Nguyen2017b,Nguyen2018}. This term requires careful consideration. Corrosion is a rate-dependent process which can be either activation-controlled or diffusion-controlled, with the role of mechanical fields in enhancing the local current density being only relevant under activation-controlled corrosion \citep{Gutman1998}. However, the last term in (\ref{eq:omega}) leads to an influence of mechanical fields in corrosion kinetics even under diffusion-controlled conditions, where corrosion is only governed by ionic transport. Hence, we drop this term and, instead, account for the interplay between metal dissolution and mechanics \textit{via} the dissipative microstress $\omega_d^{\text{dis}}$. In the application of the phase field paradigm to metallic corrosion, dissipation is governed by the interface kinetics coefficient $L$, also referred to as the mobility parameter. Accordingly, the dissipative part of the corrosion micro-stress is given by
\begin{equation}\label{eq:omega_dis}
    \begin{aligned}
    \omega_d^{\text{dis}} = \hat{b}\,\dot{\phi}_d =\frac{1}{L} \dot{\phi}_d \, .
    \end{aligned}
\end{equation}

Following the work by \citet{JMPS2021}, the mobility parameter is made dependent of the mechanical fields to incorporate two important effects: (i) the role of mechanical straining in rupturing the protective passive film, and (ii) the role of mechanical fields in enhancing corrosion kinetics. The former is accounted for in agreement with the film-rupture-dissolution-repassivation (FRDR) mechanism \citep{Parkins1987}, while the latter is introduced following \citet{Gutman1998} mechanochemical theory. Accordingly, the mechanics-enhanced mobility coefficient is defined as,
\begin{equation}\label{eq:L_cycle}
    L\left(\varepsilon^p , \sigma_h \right) =\left\{
\begin{aligned}
    k_\mathrm{m} \left(\varepsilon^p , \sigma_h \right) \, L_0 &, \,\,\,\,\,\,\,\,\text{if} \,\,\, 0 < t_i \leqslant t_0 \\
    k_\mathrm{m} \left(\varepsilon^p , \sigma_h \right) \, L_0 \ \mathrm{exp} \left(-k \left(t_i-t_0 \right)\right) &,  \,\,\,\,\,\,\,\,\text{if} \,\,\, t_0 < t_i \leqslant t_0+t_f \\
\end{aligned}
\right.
\end{equation}
\noindent where $\sigma_h$ is the hydrostatic stress, $L_0$ is the reference mobility coefficient (in the absence of mechanical straining) and $k_\mathrm{m}$ denotes the mechanochemical term. Following \citet{Gutman1998}, the mechanochemical term is defined as follows to capture the impact of mechanical fields on corrosion kinetics,
\begin{equation}\label{eq:Gutman}
  k_\mathrm{m}\left(\varepsilon^p,\sigma_h\right)=\left(\frac{\varepsilon^p}{\varepsilon_y} + 1 \right) \mathrm{exp}\left(\frac{\sigma_h V_m}{RT}\right) \, ,
\end{equation}
\noindent where $\varepsilon_y=\sigma_y/E$ is the yield strain and $V_m$ is the molar volume. In regard to the cyclic film-rupture process, as elaborated in \citet{JMPS2021}, the upper part of (\ref{eq:L_cycle}) corresponds to the behaviour when the passive film has been ruptured and thus the corrosion current density and $L$ take their highest values (bare metal corrosion). After a time $t_0$, repassivation starts to play a role and reduces the current density (and $L$). The rate of decay depends on the stability of the passive film, the material and the environment, and is characterised through the parameter $k$. The exponential shape of (\ref{eq:L_cycle})b reflects the increase in film stability with time that is observed as more oxides are deposited. After a time $t_f$ since the decay process started, a rupture event occurs and again $L=k_m L_0$. Film rupture is governed by a competition between repassivation kinetics, how long it takes for a film to stabilise, and crack tip straining, how fast the new film will accumulate strains until reaching a critical value. Accordingly, film rupture will occur when the accumulated equivalent plastic strain over a FRDR cycle $\varepsilon_i^p$ reaches a critical failure strain $\varepsilon_f$,
\begin{equation}\label{eq:tf}
    \varepsilon_i^p = \varepsilon_f \,\,\,\,\,\,\,\, \text{with} \,\,\,\,\,\,\,\, \varepsilon_i^p = \int_0^{t_i} \dot{\varepsilon}^p \, \text{d}t
\end{equation}

\noindent where the total time for each FRDR cycle equals $t_i=t_0+t_f$ and $\varepsilon_f$ is assumed to be on the order of 0.1\% \citep{Gutman2007}.\\ 

Finally, insert these constitutive relations (\ref{eq:omegadf}), (\ref{eq:zeta})-(\ref{eq:omega_dis}),  into the phase field balance equation (\ref{eq:LocalBalance}b), to re-formulate the strong form as:
\begin{equation}\label{eq:Allen–Cahn}
   \frac{1}{ L\left(\varepsilon^p,\sigma_h\right)} \frac{\mathrm{d} \phi_d}{\mathrm{d} t} +  \left( \frac{\partial \,\psi^{ch}}{\partial \phi_d}- \alpha \nabla^2 \phi_d \right)=0 \, ,
\end{equation}
\noindent with
\begin{equation}
    \begin{aligned}
   \frac{\partial \,\psi^{ch}}{\partial \phi_d} &=- 2 A \left[c_\mathrm{M} - h_d \left(\phi_d\right)(c_\mathrm{Se} - c_\mathrm{Le}) - c_\mathrm{Le} \right] (c_\mathrm{Se} - c_\mathrm{Le}) \,h_d' \left(\phi_d \right)+ w \,g ' (\phi_d)
     \end{aligned}
\end{equation}

\subsubsection{Transport of ionic species}

Let us now derive the constitutive prescriptions for the transport of metallic ions, which governs corrosion under diffusion-controlled conditions. Following (\ref{eq:constitutive})-(\ref{eq:Free_energy}), the chemical potential $\mu_\mathrm{M}$ is given by,
\begin{equation}\label{eq:mu_M}
    \mu_\mathrm{M} = \frac{1}{c_\mathrm{solid}} \frac{\partial \,\psi}{\partial c_\mathrm{M}} = \frac{2A}{c_\mathrm{solid}}\left[\left(c_\mathrm{M} - h_d\left(\phi_d\right)(c_\mathrm{Se} - c_\mathrm{Le}) - c_\mathrm{Le} \right)\right]
\end{equation}

While the flux $\mathbf{J}_\mathrm{M}$ can be determined following a Fick law-type relation as,
\begin{equation}\label{eq:flux_M}
    \mathbf{J}_\mathrm{M} = - \frac{D_\mathrm{M}}{2A} \cdot c_\mathrm{solid} \cdot c_\mathrm{solid} \cdot \nabla \mu_\mathrm{M} = c_\mathrm{solid} \cdot D_\mathrm{M} \nabla \left[\left(c_\mathrm{M} - h_d\left(\phi_d\right)(c_\mathrm{Se} - c_\mathrm{Le}) - c_\mathrm{Le} \right)\right]  
\end{equation}
\noindent where $D_\mathrm{M}$ is the diffusion coefficient of metal ion. Finally, inserting (\ref{eq:mu_M}) and (\ref{eq:flux_M}) into the mass transport balance (\ref{eq:LocalBalance}c), one can subsequently obtain the following governing equation for metal ion diffusion:
\begin{equation}\label{eq:Cahn–Hilliard}
    \begin{aligned}
     \frac{\mathrm{d} c_\mathrm{M}}{\mathrm{d} t} - \nabla \cdot D_\mathrm{M} \nabla \left[\left(c_\mathrm{M} - h_d \left(\phi_d\right)(c_\mathrm{Se} - c_\mathrm{Le}) - c_\mathrm{Le} \right)\right] = 0 
    \end{aligned}
\end{equation}

\subsubsection{A phase field description of fracture}

We now turn our attention to the constitutive relations for phase field fracture. The energetic microstress $\omega_f^{\text{en}}$ in (\ref{eq:constitutive}) is given by:
\begin{equation}\label{eq:Omega_f}
    \omega_f^{\text{en}} = \dfrac{\partial\,\psi}{\partial\phi_f} = h_f'\left(\phi_f\right)\, h_d\left(\phi_d\right)\, \psi_0^m -  G_c \left(c_\mathrm{H}\right) \frac{1-\phi_f}{\ell_f}
\end{equation}

Here, fracture is assumed to be a rate-independent process such that $\hat{\beta}=0$ in (\ref{eq:Omega_fdis}). In addition, the phase field microstress vector $\mathbf{\zeta}_f$ reads:
\begin{equation}\label{eq:zeta_f}
    \mathbf{\zeta}_f = \dfrac{\partial\,\psi}{\partial \nabla \phi_f} =  G_c \left(c_\mathrm{H}\right) \, \ell_f \, \nabla \phi_f
\end{equation}

Inserting (\ref{eq:Omega_f})-(\ref{eq:zeta_f}) into the phase field local balance (\ref{eq:LocalBalance}d), the phase field fracture balance equation can be formulated as:
\begin{equation}\label{eq:PhaseField_f}
  h_f'\left(\phi_f\right)\, h_d\left(\phi_d\right)\, \psi_0^m +  G_c \left(c_\mathrm{H}\right) \left( -\frac{1-\phi_f}{\ell_f} - \ell_f {\nabla^2} \phi_f \right) = 0  \, ,
\end{equation}

\noindent where, as in \cite{CMAME2018}, we have assumed that the concentration gradient is small at the interface.

\subsubsection{Diffusion of atomic hydrogen}

We conclude the description of our constitutive theory by deriving the appropriate relations for hydrogen transport. In agreement with (\ref{eq:constitutive})-(\ref{eq:Free_energy}), the chemical potential is given by,
\begin{equation}\label{eq:mu_H}
    \mu_\mathrm{H}= \frac{\partial\,\psi}{\partial c_\mathrm{H}} = \mu_0 + RT\ln{\dfrac{\theta_\mathrm{H}}{1-\theta_\mathrm{H}}} - \overline{V}_\mathrm{H} \sigma_h  + \frac{\mathrm{d} \,  G_c \left(c_\mathrm{H}\right)}{\mathrm{d} \, \theta} \frac{\mathrm{d} \, \theta }{\mathrm{d} c_\mathrm{H}} \left[ \frac{\ell_f}{2} \, |\nabla \phi_f|^2 + \frac{1}{2\,\ell_f} \left(1-\phi_f\right)^2 \right]
\end{equation}

The last term in (\ref{eq:mu_H}), which enhances hydrogen transport from damaged regions to pristine regions, is neglected as a penalty-based \emph{moving} chemical boundary condition is implemented to capture how the aqueous solution immediately occupies the space created by crack advance \citep{CS2020}. Accordingly, the hydrogen flux $\mathbf{J}_\mathrm{H}$ is determined through a linear Onsager relationship,
\begin{equation}
    \mathbf{J}_\mathrm{H}= - \frac{D_\mathrm{H} c_\mathrm{H}}{RT} \nabla \mu_\mathrm{H} = - \frac{D_\mathrm{H} c_\mathrm{H}}{\theta_\mathrm{H}\left(1-\theta_\mathrm{H}\right)} \nabla \theta_\mathrm{H} + \frac{D_\mathrm{H} c_\mathrm{H}}{RT} \overline{V}_\mathrm{H} \nabla \sigma_h \,
\end{equation}
\noindent which, under the common assumption of low occupancy $\left(\theta_\mathrm{H} \ll 1\right)$, reads
\begin{equation}\label{eq:flux_H}
    \mathbf{J}_\mathrm{H}= - D_\mathrm{H} \nabla c_\mathrm{H} + \frac{D_\mathrm{H} c_\mathrm{H}}{RT} \overline{V}_\mathrm{H} \nabla \sigma_h
\end{equation}

Inserting (\ref{eq:mu_H}) and (\ref{eq:flux_H}) into the hydrogen transport balance (\ref{eq:LocalBalance}e), the strong form of the atomic hydrogen diffusion problem becomes:
\begin{equation}\label{eq:hydrogen transport}
    \begin{aligned}
     \frac{\mathrm{d} c_\mathrm{H}}{\mathrm{d} t} - D_\mathrm{H} \nabla c_\mathrm{H} + \frac{D_\mathrm{H} c_\mathrm{H}}{RT} \overline{V}_\mathrm{H} \nabla \sigma_h = 0 
    \end{aligned}
\end{equation}

\subsection{Summary of the governing equations}
\label{Sec:SummaryTheory}

\subsubsection{Strong form and multi-physics interactions}

Our formulation resolves five key physical processes, namely mechanical deformation, metal dissolution, ionic transport, fracture, and hydrogen diffusion. We shall now summarise the governing equations obtained for the associated primal field quantities ($\mathbf{u},\,\phi_d,\,c_\mathrm{M}, \,\phi_f,\,c_\mathrm{H})$, see Table \ref{tab:strongform}, and discuss their interactions.

\begin{table} [!htb]
  \raggedleft
  \caption{Summary of the strong form of the governing equations of the model.}
    \begin{tabular*}{\hsize}{@{}@{\extracolsep{\fill}}lr@{}}
    \toprule
    \textbf{Governing equations:} \vspace{4.0mm} \\ 
     $\qquad \nabla \cdot \bm{\sigma} = \mathbf{0}$ & (T.1)\,\,\,\,\,\,\,\vspace{2.5mm} \\ 
     $\qquad \frac{1}{ L\left(\varepsilon^p,\sigma_h\right)} \frac{\mathrm{d} \phi_d}{\mathrm{d} t} + \left( \frac{\partial \,\psi^{ch}}{\partial \phi_d}- \alpha \nabla^2 \phi_d \right)=0$ & (T.2)\,\,\,\,\,\,\,\vspace{2.5mm} \\
     $\qquad \frac{\mathrm{d} c_\mathrm{M}}{\mathrm{d} t} - \nabla \cdot D_\mathrm{M} \nabla \left[\left(c_\mathrm{M} - h_d\left(\phi_d\right)(c_\mathrm{Se} - c_\mathrm{Le}) - c_\mathrm{Le} \right)\right] = 0 $ & (T.3)\,\,\,\,\,\,\,\vspace{2.5mm} \\
     $\qquad h_f'\left(\phi_f\right)\, h_d\left(\phi_d\right)\, \psi_0^m +  G_c \left(c_\mathrm{H}\right) \left( -\frac{1-\phi_f}{\ell_f} - \ell_f {\nabla^2} \phi_f \right) = 0 $ & (T.4)\,\,\,\,\,\,\,\vspace{2.5mm} \\     
     $\qquad \frac{\mathrm{d} c_\mathrm{H}}{\mathrm{d} t} - D_\mathrm{H} \nabla c_\mathrm{H} + \frac{D_\mathrm{H} c_\mathrm{H}}{RT} \overline{V}_\mathrm{H} \nabla \sigma_h = 0 $ & (T.5)\,\,\,\,\,\,\,\vspace{2.5mm} \\
     
     with: \vspace{4.0mm} \\ 
     $\qquad \bm{\sigma} = \frac{\partial \, \psi}{\partial \bm{\varepsilon}} = h_d\left(\phi_d \right) h_f\left(\phi_f \right)\bm{C^{\mathrm{ep}}}:\mathrm{sym} \nabla  \mathbf{u} $ & (T.6)\,\,\,\,\,\,\,\vspace{2.5mm} \\ 
     $\qquad \frac{\partial \,\psi^{ch}}{\partial \phi_d}=- 2 A \left[c_\mathrm{M} - h_d\left(\phi_d\right)(c_\mathrm{Se} - c_\mathrm{Le}) - c_\mathrm{Le} \right] (c_\mathrm{Se} - c_\mathrm{Le}) \,h_d'\left(\phi_d \right) + w \,g ' (\phi_d) $ & (T.7)\,\,\,\,\,\,\,\vspace{2.5mm} \\ 
     $\qquad \psi_0^m = \frac{1}{2} \bm{\varepsilon}^e : \bm{C}^e : \bm{\varepsilon}^e + \int_0^t \bm{\sigma}_0 : \dot{\bm{\varepsilon}}^p \, \text{d}t $ & (T.8)\,\,\,\,\,\,\, \\     
    \bottomrule
    \end{tabular*}
  \label{tab:strongform}
\end{table}

These governing equations account for multiple coupling phenomena, as described below.

\begin{itemize}[leftmargin=*]
    \item Metal dissolution and ionic transport\\
    When diffusion-controlled corrosion conditions are relevant (i.e., a sufficiently large $L$), the evolution of $\phi_d$ is driven by the concentration of metal ions - see the first term on the right hand side of (T.7). Also, as shown in (T.3), the dissolution phase field interacts with the ionic transport equation; the diffusion of metal ions is restricted by $\phi_d$ such that it only takes place along the corrosion interface and in the electrolyte.
    
    \item Metal dissolution and mechanical fields\\
    The dissolution process redistributes mechanical fields by reducing the stiffness of the solid through a degradation function $h_d\left(\phi_d \right)$, see (T.6). Additionally, the mechanical fields impact the corrosion behaviour in two ways: (i) by enhancing corrosion kinetics, and (ii) by breaking the protective passive film. Both effects are incorporated by defining the mobility coefficient dependent on the relevant mechanical quantities, $L=L\left(\varepsilon^p,\sigma_h\right)$ - see (T.2) and (\ref{eq:L_cycle}).
    
    \item Metal dissolution and fracture\\
    The interaction of metal dissolution and physical fracture is implemented by defining a joint degradation term $ h_d\left(\phi_d \right) \cdot h_f\left(\phi_f \right)$, see (T.6). As a result, metal dissolution provides an additional term to the conventional phase field fracture balance equation, preventing cracks from nucleating in the electrolyte, see (T.4). On the other hand, no action is taken to prevent corrosion from happening in cracked regions due to the very different time scales involved and the fact that corrosion typically precedes structural failure. For visualisation purposes, an equivalent damage variable $\phi_e$ is defined to describe the joint effect of dissolution and fracture damage: $\phi_e=\phi_d\cdot\phi_f$.
    
    \item Mechanical fields and hydrogen diffusion\\
    Mechanical fields and hydrogen transport are coupled in one direction as higher hydrostatic stresses (volumetric strains) drive hydrogen transport due to lattice dilation; this is accounted for by the last term in (T.5). Thus, hydrogen accumulates ahead of pits, cracks and other defects.
    
    \item Fracture and hydrogen diffusion\\
    There is a two-way interaction between hydrogen diffusion and fracture. First, the material toughness is defined to be a function of the hydrogen content $G_c \left(c_\mathrm{H}\right)$, see (T.4). This captures the embrittlement effect of hydrogen. Second, as detailed below, a penalty approach is used to enforce $c_\mathrm{H}=c_\mathrm{env}$ in cracked regions, where $c_\mathrm{env}$ is the hydrogen content associated with the environment.
    
\end{itemize}

\subsubsection{A brief discussion on $\ell_d$ and $\ell_f$}
\label{Sec:ldlf}

The proposed multi-phase-field formulation incorporates two length scale parameters, $\ell_d$ for the dissolution process and $\ell_f$ for the fracture process. Both provide a measure of the interface thickness. In the case of phase field corrosion, $\ell_d$ is not an explicit parameter but a byproduct of the height of the double well potential $w$ and the gradient energy coefficient $\alpha$ - see (\ref{eq:gammal}). The interface energy $\gamma$ and the thickness of the reaction interface $\ell_d$ can, in principle, be estimated from experiments \citep{Abubakar2015} and, consequently, $w$ and $\alpha$ can be determined as follows:
\begin{equation}\label{eq:gammal2}
    \begin{aligned}
    w = \frac{6\,\gamma a^* }{\ell_d}   \,\,\,\,\,\,\,\, \text{and} \,\,\,\,\,\,\,\, \alpha = \frac{3\,\gamma \ell_d}{a^*} \, .
    \end{aligned}
\end{equation}

Regarding the phase field fracture length scale, $\ell_f$; it can be seen as a regularising parameter in the limit $\ell_f \to 0^+$ but it is generally considered to be a material parameter that defines the material strength $\sigma_c$ \citep{Tanne2018,PTRSA2021}. This can be readily observed by considering the homogeneous solution to (\ref{eq:PhaseField_f}), giving a maximum stress of:
\begin{equation}\label{eq:sigma_c}
    \sigma_c = \frac{3}{16} \sqrt{\frac{3 E G_c }{\ell_f}} \, .
\end{equation}

\section{Numerical implementation}
\label{Sec:Numerical implementation}

We proceed to briefly describe the details of the numerical implementation, including the formulation of the weak form (Section \ref{Sec:WeakForm}), the finite element discretisation (Section \ref{Sec:FEdiscretisation}) and the derivation of the residuals and stiffness matrices (Section \ref{Sec:ResidualsStiffnessMatrices}). 

\subsection{Weak form}
\label{Sec:WeakForm}

The weak form of the balance equations in Table \ref{tab:strongform} can be readily derived as:
\begin{flalign}\label{eq:Weak_u}
    \int_\Omega \left[h_d\left(\phi_d\right) h_f\left(\phi_f\right)+\kappa \right] \bm{\sigma}_0 \delta \bm{\varepsilon}\,\mathrm{d}V - \int_{\partial \Omega} \mathbf{T} \cdot \delta \mathbf{u}\,\mathrm{d}S = 0 &&
\end{flalign} \vspace{-0.6cm}
\begin{flalign}\label{eq:Weak_phid}
   \int_\Omega  \frac{1}{ L\left(\varepsilon^p,\sigma_h\right)} \frac{\mathrm{d} \phi_d}{\mathrm{d} t} \delta \phi_d \,\mathrm{d}V + \int_\Omega \frac{\partial \,\psi^{ch}}{\partial \phi_d} \delta \phi_d \,\mathrm{d}V +  \int_\Omega \alpha \nabla \phi_d \cdot \nabla \delta \phi_d \,\mathrm{d}V = 0 &&
\end{flalign} \vspace{-0.8cm}
\begin{flalign}\label{eq:Weak_cM}
    \int_\Omega \frac{1}{D_\mathrm{M}} \frac{\mathrm{d} c_\mathrm{M}}{\mathrm{d} t} \delta c_\mathrm{M} \,\mathrm{d}V + \int_\Omega \nabla \left[c_\mathrm{M} - h_d\left(\phi_d\right)(c_\mathrm{Se} - c_\mathrm{Le}) - c_\mathrm{Le} \right] \cdot \nabla \delta c_\mathrm{M} \, \mathrm{d}V = 0 &&
\end{flalign} \vspace{-0.8cm}
\begin{flalign}\label{eq:Weak_phif}
    \int_\Omega   h_f'\left(\phi_f\right)\, h_d\left(\phi_d\right)\, \mathcal{H} \,\mathrm{d}V + \int_\Omega  G_c \left(c_\mathrm{H}\right) \left( -\frac{1-\phi_f}{\ell_f} \delta \phi_f + \ell_f \nabla \phi_f \cdot \nabla \delta \phi_f \right) \, \mathrm{d}V = 0 &&
\end{flalign} \vspace{-0.8cm}
\begin{flalign}\label{eq:Weak_cH}
    \int_\Omega \frac{1}{D_\mathrm{H}} \frac{\mathrm{d} c_\mathrm{H}}{\mathrm{d} t} \delta c_\mathrm{H} \,\mathrm{d}V + \int_\Omega \nabla c_\mathrm{H} \cdot \nabla \delta c_\mathrm{H}\, \mathrm{d}V - \int_\Omega \frac{\overline{V}_\mathrm{H} c_\mathrm{H}}{RT} \nabla \sigma_h \cdot \nabla \delta c_\mathrm{H} \, \mathrm{d}V = 0 &&
\end{flalign}    
\noindent where $\kappa=1 \times 10^{-5}$ is a parameter chosen to keep the system of equations well-conditioned and $\mathcal{H}$ is the driving force for fracture, which incorporates a history variable field to ensure damage irreversibility \citep{Miehe2010a}. Since the effective plastic work increases monotonically, the irreversibility condition is only applied to the elastic part. Thus, in its most general form, the driving force for fracture is given by,
\begin{equation}\label{eq:Weightingfactors}
    \begin{aligned}
     \mathcal{H}= \beta_e \mathcal{H}^e + \beta_p \psi_0^{p} = \beta_e \mathop{\mathrm{max}}\limits_{\tau \in \left[0,t \right]} \psi_0^{e+} \left(\tau \right) + \beta_p \psi_0^{p} \, .
    \end{aligned}
\end{equation}

\noindent where $\beta_e$ and $\beta_p$ are weighting factors \citep{Borden2016}, which are taken to be equal to 1, unless otherwise stated. Also, $\psi_0^{e+}$ refers to the tensile part of the elastic free energy density; $\psi_0^e=\psi_0^{e+}+\psi_0^{e-}$. This decomposition is applied to prevent damage under compression, and we choose to adopt the so-called spherical/deviatoric split \citep{Amor2009},
\begin{equation}
    \psi_0^{e+}=\frac{1}{2} K \langle \text{tr} \left(\bm{\varepsilon}^e\right)\rangle_+^2 + \mu \left( \bm{\varepsilon}^e{'}: \bm{\varepsilon}^e{'} \right) \,\,\,\,\,\,\,\, \text{and} \,\,\,\,\,\,\,
    \psi_0^{e-}=\frac{1}{2}K\langle \text{tr} \left(\bm{\varepsilon}^e\right)\rangle_-^2 
\end{equation}

\noindent where $\left\langle \Hsquare \right\rangle$ denote Macaulay brackets and $\mu$ is the shear modulus.

\subsection{Finite element discretisation}
\label{Sec:FEdiscretisation}

The finite element method is used to discretise and solve the coupled equations (\ref{eq:Weak_u})-(\ref{eq:Weak_cH}). Using Voigt's notation, the nodal variables for the displacement field $\hat{\mathbf{u}}$, the phase field corrosion order parameter $\hat{\phi}_d$, the normalised metal ion concentration $\hat{c}_\mathrm{M}$, the phase field fracture order parameter $\hat{\phi}_f$, and the hydrogen concentration $\hat{c}_\mathrm{H}$ are interpolated as:
\begin{equation}
    \begin{aligned}\label{eq:FEdis1}
    &\mathbf{u} = \sum_{i=1}^m \bm{N}_\emph{i}^{\mathbf{u}}\, \mathbf{\hat{u}_\emph{i}}, \,\,\, \phi_d = \sum_{i=1}^m N_i \,\hat{\phi}_{d_i}, \,\,\, c_\mathrm{M} = \sum_{i=1}^m N_i\, \hat{c}_{\mathrm{M}_i}, \,\,\, \phi_f = \sum_{i=1}^m N_i\, \hat{\phi}_{f_i}, \,\,\, c_\mathrm{H} = \sum_{i=1}^m N_i \,\hat{c}_{\mathrm{H}_i}
    \end{aligned}
\end{equation}

\noindent where $N_i$ denotes the shape function associated with node $i$, for a total number of nodes $m$, and $\bm{N}_\emph{i}^{\mathbf{u}}$ is a diagonal interpolation matrix with the nodal shape functions $N_i$ as components. Similarly, using the standard strain-displacement $\bm{B}$ matrices, the associated gradient quantities are discretised as: 
\begin{equation}
    \begin{aligned}\label{eq:FEdis2}
    &\bm{\varepsilon} = \sum_{i=1}^m \bm{B}_\emph{i}^{\mathbf{u}} \,\mathbf{\hat{u}_\emph{i}},\,\,\, \nabla \phi_d = \sum_{i=1}^m \mathbf{B}_\emph{i} \,\hat{\phi}_{d_i},\,\,\, \nabla c_\mathrm{M} = \sum_{i=1}^m \mathbf{B}_\emph{i} \,\hat{c}_{\mathrm{M}_i}, \,\,\, \nabla \phi_f = \sum_{i=1}^m \mathbf{B}_\emph{i} \,\hat{\phi}_{f_i},\,\,\, \nabla c_\mathrm{H} = \sum_{i=1}^m \mathbf{B}_\emph{i}\, \hat{c}_{\mathrm{H}_i}
    \end{aligned}
\end{equation}

\subsection{Residuals and stiffness matrices}
\label{Sec:ResidualsStiffnessMatrices}

Considering the finite element discretisation (\ref{eq:FEdis1})-(\ref{eq:FEdis2}), the weak form equations (\ref{eq:Weak_u})-(\ref{eq:Weak_cH}) can be discretised to derive the following residuals,
\begin{flalign}\label{r_u}
   \mathbf{r}_{i}^{\mathbf{u}} = \int_\Omega \left[ h_d\left(\phi_d\right) h_f\left(\phi_f\right) + \kappa \right] (\bm{B}_\emph{i}^{\mathbf{u}})^T \bm{\sigma}_0 \, \mathrm{d}V - \int_{\partial \Omega} (\bm{N}_\emph{i}^{\mathbf{u}})^T \mathbf{T} \, \mathrm{d}S  &&
\end{flalign} \vspace{-0.8cm}
\begin{flalign}\label{r_phid}
    r_{i}^{\phi_d} = \int_\Omega  \frac{1}{ L\left(\varepsilon^p,\sigma_h\right)} \frac{\mathrm{d} \phi_d}{\mathrm{d\emph{t}}} N_i \,\mathrm{d}V + \int_\Omega \frac{\partial \psi^{ch}}{\partial \phi_d} N_i \,\mathrm{d}V + \int_\Omega \alpha \mathbf{B}_\emph{i}^T \nabla \phi_d \,\mathrm{d}V &&
\end{flalign} \vspace{-0.6cm}
\begin{flalign}\label{r_cM}
    r_{i}^{c_\mathrm{M}} = \int_\Omega \frac{1}{D_\mathrm{M}} \frac{\mathrm{d} c_\mathrm{M}}{\mathrm{d\emph{t}}} N_i \,\mathrm{d}V + \int_\Omega \mathbf{B}_\emph{i}^T \left[\nabla c_\mathrm{M} - h_d '  (\phi_d)(c_\mathrm{Se} - c_\mathrm{Le}) \nabla \phi_d \right] \, \mathrm{d}V &&
\end{flalign}
\begin{flalign}\label{r_phif}
    r_{i}^{\phi_f} = \int_\Omega h_f'\left(\phi_f\right)\, h_d\left(\phi_d\right) N_i \, \mathcal{H} \, \mathrm{d}V +  G_c \left(c_\mathrm{H}\right) \int_\Omega \left( -\frac{1-\phi_f}{\ell_f} N_i + \ell_f \mathbf{B}_\emph{i}^T \nabla \phi_f \right) \, \mathrm{d}V &&
\end{flalign} \vspace{-0.6cm}
\begin{flalign}\label{r_H}
    r_{i}^{c_\mathrm{H}} &= \int_\Omega \left[\frac{1}{D_\mathrm{H}} \frac{\mathrm{d} c_\mathrm{H}}{\mathrm{d\emph{t}}} + \left( c_\mathrm{H}-c_\mathrm{env}\right) \left\langle 1-2\phi_d  \phi_f \right\rangle k_p \right]N_i \,\mathrm{d}V + \int_\Omega \mathbf{B}_\emph{i}^T \nabla c_\mathrm{H} \, \mathrm{d}V  - \int_\Omega \mathbf{B}_\emph{i}^T \left(\frac{\overline{V}_\mathrm{H} c_\mathrm{H}}{RT} \right)\nabla \sigma_h \, \mathrm{d}V &&
\end{flalign}

\noindent where $k_p$ is a penalty factor to enforce $c_\mathrm{H}=c_\mathrm{env}$ when $\phi_d  \phi_f $ approaches zero, the moving chemical boundary conditions that mimic the prompt exposure of newly created defect surfaces to the environment \citep{CS2020}. Specifically, we find that a linear transition from $\phi_d  \phi_f =0.5$ and a $k_p$ value of $10^{5}$ adequately enforce $c_\mathrm{H}=c_\mathrm{env}$ while providing good numerical stability. Accordingly, the tangent stiffness matrices are then calculated as:
\begin{flalign}\label{K_u}
    \bm{K}_{ij}^{\mathbf{u}} = \int_\Omega \left[ h_d\left(\phi_d\right) h_f\left(\phi_f\right) + \kappa \right] (\bm{B}_\emph{i}^{\mathbf{u}})^T \bm{C_{\mathrm{ep}}} \,\bm{B}_\emph{j}^{\mathbf{u}} \, \mathrm{d}V &&
\end{flalign}\vspace{-0.8cm}
\begin{flalign}\label{K_phid}
    \bm{K}_{ij}^{\phi_d} =\int_\Omega  \frac{1}{ L\left(\varepsilon^p,\sigma_h\right)}  \frac{N_\emph{i}^T N_\emph{j} }{\mathrm{d\emph{t}}} \, \mathrm{d}V + \int_\Omega \frac{\partial ^2  \psi^{ch}}{\partial {\phi_d}^{2}} N_\emph{i}^T N_j\,\mathrm{d}V + \int_\Omega \alpha \mathbf{B}_\emph{i}^T \mathbf{B}_\emph{j} \,\mathrm{d}V &&
\end{flalign}\vspace{-0.8cm}
\begin{flalign}\label{K_cM}
    \bm{K}_{ij}^{c_\mathrm{M}} = \int_\Omega \frac{1}{D_\mathrm{M}} \frac{N_\emph{i}^T N_\emph{j} }{\mathrm{d\emph{t}}} \, \mathrm{d}V + D_\mathrm{M} \int_\Omega \mathbf{B}_\emph{i}^T \mathbf{B}_\emph{j} \, \mathrm{d}V &&
\end{flalign}\vspace{-0.6cm}
\begin{flalign}\label{K_phif}
    \bm{K}_{ij}^{\phi_f} = \int_\Omega \left\{\left[ h_f''(\phi_f)\, h_d\left(\phi_d\right) \, \mathcal{H} + \frac{ G_c \left(c_\mathrm{H}\right)}{\ell_f}\right] N_\emph{i}^T N_j +  G_c \left(c_\mathrm{H}\right) \ell_f \mathbf{B}_\emph{i}^T \mathbf{B}_\emph{j}\right\} \,\mathrm{d}V &&
\end{flalign}\vspace{-0.6cm}
\begin{flalign}\label{K_cH}
    \bm{K}_{ij}^{c_\mathrm{H}} = \int_\Omega \frac{1}{D_\mathrm{H}} \frac{N_\emph{i}^T N_\emph{j}}{\mathrm{d\emph{t}}} \, \mathrm{d}V + \int_\Omega N_\emph{i}^T N_\emph{j} \left\langle 1-2\phi_d  \phi_f \right\rangle k_p \, \mathrm{d}V + \int_\Omega \mathbf{B}_\emph{i}^T \mathbf{B}_\emph{j} \, \mathrm{d}V - \int_\Omega \mathbf{B}_\emph{i}^T \left(\frac{\overline{V}_\mathrm{H}}{RT} \nabla \sigma_h \right) N_j \, \mathrm{d}V &&
\end{flalign}

The resulting finite element system is solved using time parametrisation and an incremental-iterative Newton-Raphson scheme. The implementation is carried out in the commercial finite element package \texttt{ABAQUS} by means of a user element (\texttt{UEL}) subroutine. Abaqus2Matlab \citep{AES2017} is used for pre-processing the input files. Following \citet{CMAME2018}, $\nabla \sigma_h$ is computed utilising the derivatives of the shape functions and the hydrostatic stress values at the integration points.

\section{Results}
\label{Sec:Results}

By means of several case studies of particular interest, we proceed to showcase the potential of the model in predicting SCC driven by dissolution mechanisms, hydrogen embrittlement or a combination of both. Firstly, we simulate the growth of a semi-circular pit (Section \ref{Sec:semicorro}) and the failure of an embrittled notched plate (Section \ref{Sec:SENThydrogen}), two paradigmatic benchmarks that serve to validate the model. Secondly, we investigate the SCC failures of stainless steel samples under pure bending, considering the role of the passivation film in preventing hydrogen uptake and the transition from anodic-driven SCC to hydrogen-driven SCC (Section \ref{Sec:bending test}). Thirdly, in Section \ref{Sec:pit-to-crack transition}, cracking thresholds are predicted for a wide range of environments, spanning the regimes of dominance of hydrogen embrittlement and anodic dissolution, as well as the transition between them. Numerical results are compared with experimental measurements in artificial and biologically-active seawater. Finally, we model the seminal experiments by \citet{Gruhl1984} in Section \ref{Sec:GruhlExpt} to gain insight into the competition between hydrogen embrittlement and anodic dissolution damage mechanisms.

\subsection{Growth of a semi-circular pit}
\label{Sec:semicorro}

\begin{figure}[H]
\centering
\noindent\makebox[\textwidth]{%
\includegraphics[scale=0.42]{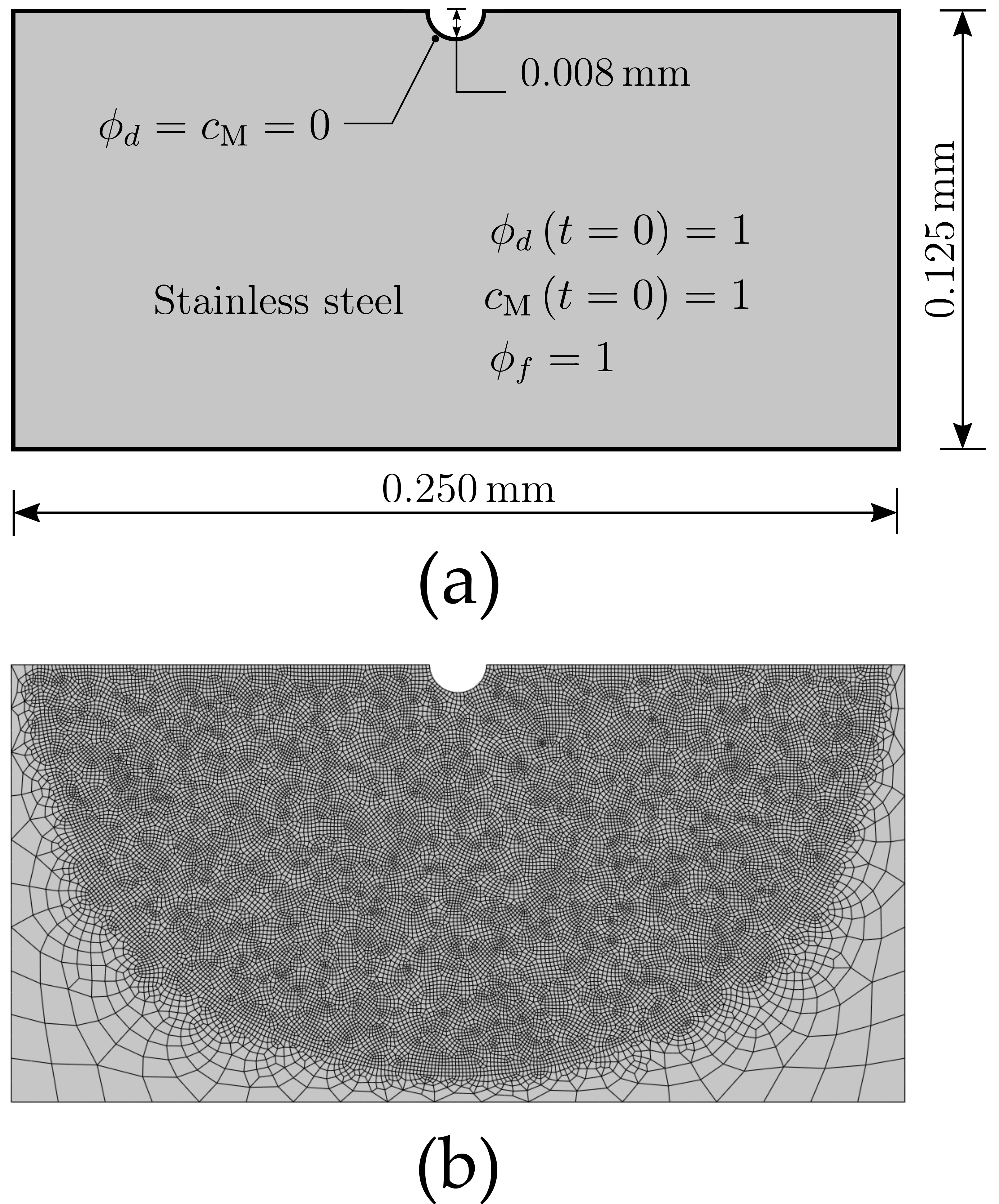}}
\caption{Growth of a semi-circular pit: (a) geometric setup, with the initial and boundary conditions, and (b) finite element mesh.}
\label{fig:semicircle}
\end{figure}

We shall begin by validating model predictions under pure corrosion conditions. Specifically, we simulate pitting corrosion in the absence of mechanical forces. As it has been shown experimentally \citep{Ernst2002}, a single growing pit will maintain a semi-circular shape throughout the whole corrosion process if corrosion is diffusion-controlled and only a small region in the edge of the metal is exposed to the electrolyte. To numerically reproduce these observations, we consider a 0.250 mm $\times$ 0.125 mm rectangular domain with a small semi-circular opening of radius 0.008 mm. The geometric setup, dimensions and boundary conditions are detailed in Fig. \ref{fig:semicircle}a. The material and corrosion parameters employed are listed in Table \ref{tab:corropara}. These are chosen to mimic previous computational studies of this boundary value problem, using either the level set method \citep{Duddu2014} or phase field formulations \citep{Mai2016,Gao2020}. A sufficiently large magnitude of $L\equiv L_0$ is adopted to ensure diffusion-controlled corrosion conditions. In this case study, pitting corrosion occurs in isolation from other phenomena and thus no mechanical load, hydrogen transport or mechanical damage are considered. As shown in Fig. \ref{fig:semicircle}b, the finite element mesh is refined in the expected corrosion region. The characteristic element size is 0.001 mm, five times smaller than the interface thickness $\ell_d$, and approximately 25,000 plane strain quadratic quadrilateral elements are used.  

\begin{figure}[H]
\centering
\noindent\makebox[\textwidth]{%
\includegraphics[scale=0.3]{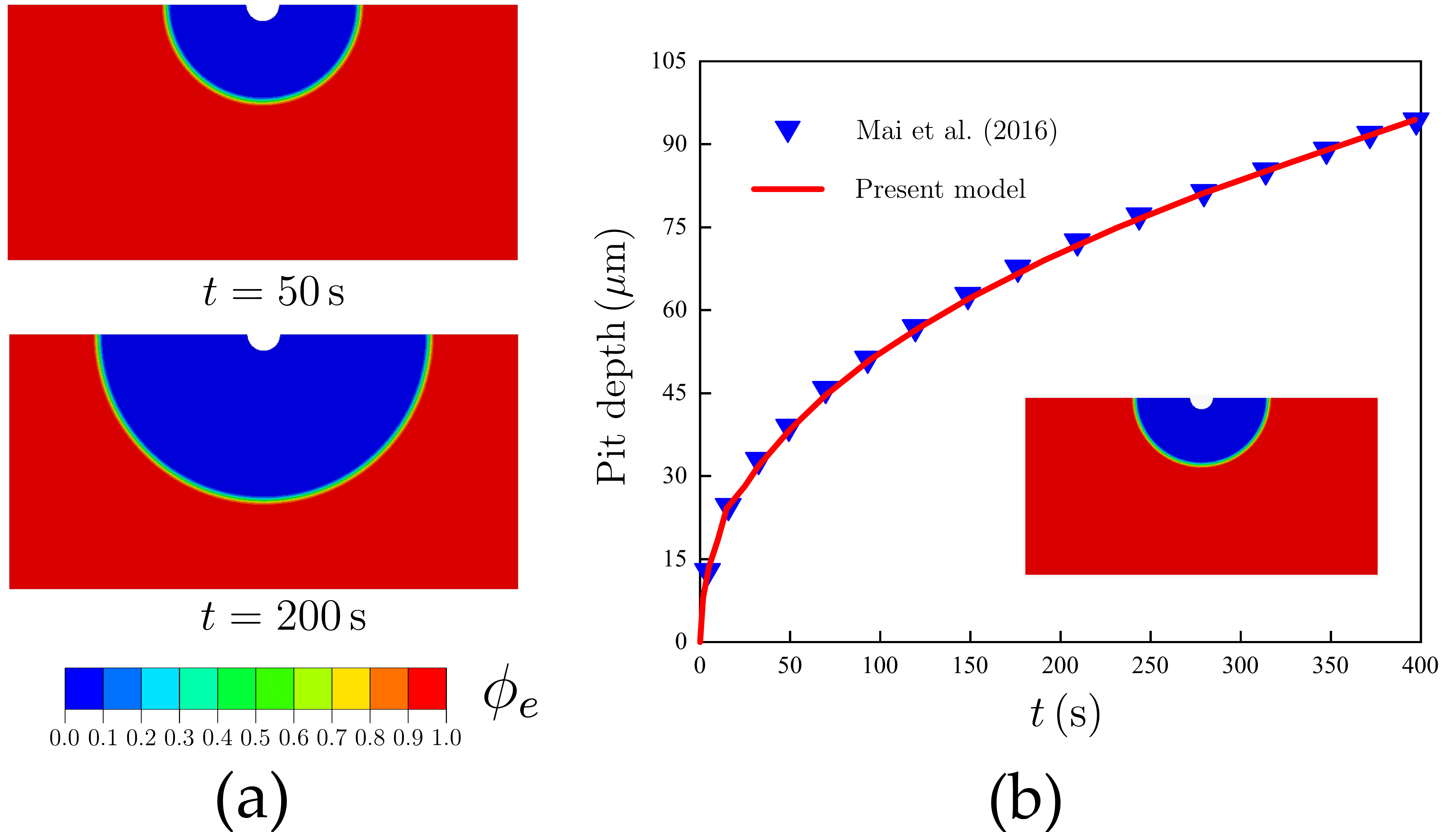}}
\caption{Growth of a semi-circular pit: (a) contours of the damage variable $\phi_e$, and (b) predictions of pit depth versus time, including the results by \citet{Mai2016}.}
\label{fig:PC1}
\end{figure}

The results obtained are shown in Fig. \ref{fig:PC1}. The evolution of the corrosion front is captured by means of the equivalent phases field $\phi_e$, which is equal to the dissolution phase field in this case study. In agreement with experiments and previous theoretical studies, the pit grows uniformly with time (see Fig. \ref{fig:PC1}a), with the pit depth versus time exhibiting a parabolic trend typical of diffusion-controlled corrosion (see Fig. \ref{fig:PC1}b). An excellent agreement is attained with the results by \citet{Mai2016}, validating the pitting corrosion predictive capabilities of our model and the numerical implementation. The contours of $c_\text{M}$ (not shown) mimic those of $\phi_e$, indicating that the diffusion of metal ions occurs only on the interface and in the electrolyte, in agreement with expectations - see (T.3).

\begin{table} [H]
  \centering
  \caption{Electrochemical parameters for the semi-circular pitting growth test.}
    \begin{tabular*}{\hsize}{@{}@{\extracolsep{\fill}}lll@{}}
    \toprule
    \textbf{Parameter} & \textbf{Value} & {\textbf{Unit}} \\
    \midrule
    Interface energy $\gamma$ & 10  & $\text{J}/\text{m}^2$ \\
    Interface thickness $\ell_d$ & 0.005  & $\mathrm{mm}$ \\
    Temperature $T$ & 300  & $\mathrm{K}$ \\
    Diffusion coefficient of metal ion $D_\text{M}$ & $8.5\times10^{-4}$ & $\mathrm{mm^2/s}$ \\
    Interface kinetics coefficient $L_0$ & $2\times10^6$ & $\mathrm{mm^2/(N \cdot s)}$ \\
    Free energy density curvature $A$ & 53.5 & $\mathrm{N/mm^2}$ \\
    Average concentration of metal $c_\mathrm{solid}$ & 143 & $\mathrm{mol/L}$ \\
    Average saturation concentration $c_\mathrm{sat}$ & 5.1 & $\mathrm{mol/L}$ \\
    \bottomrule
    \end{tabular*}
  \label{tab:corropara}
\end{table}

\begin{figure}[H]
\centering
\noindent\makebox[\textwidth]{%
\includegraphics[scale=0.37]{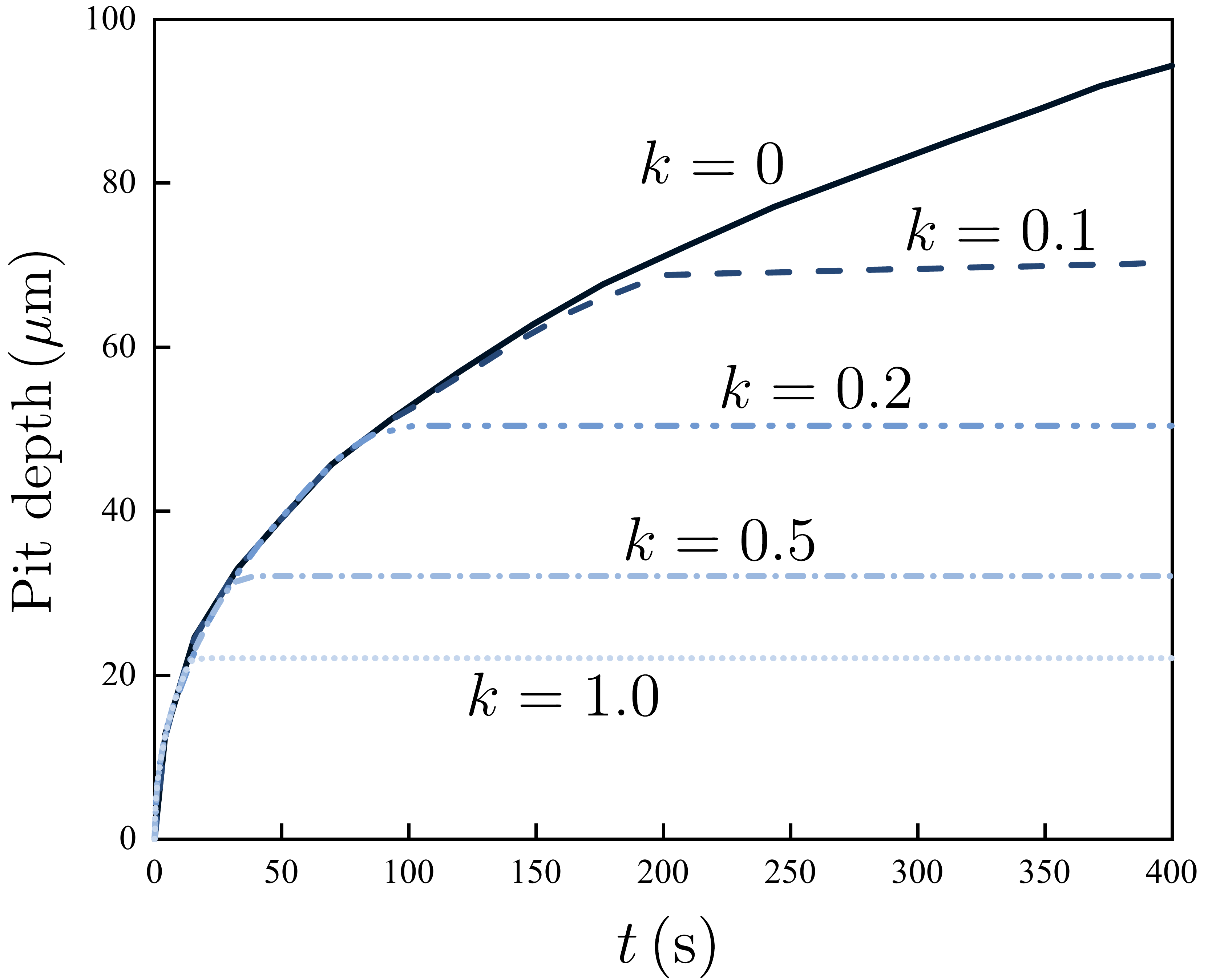}}
\caption{Growth of a semi-circular pit: Pit growth as a function of corrosion time for five selected values of the film stability parameter $k$.}
\label{fig:Film}
\end{figure}

Now, we extend this case study to investigate the role of the passive film. Thus, in addition to the case of $k=0$ (no film), shown in Fig. \ref{fig:PC1}, we consider the presence of a passivation film with different degrees of stability and protection, as characterised by selected values of $k$ (namely, 0.1, 0.2, 0.5 and 1). The time interval $t_0$ in (\ref{eq:L_cycle}) is chosen to be 0, implying that the film starts protecting the surface from its creation. Also, we emphasise that no mechanical loading is applied in this study, so the film never ruptures and just increases its protective capabilities with time. The results obtained are shown in Fig. \ref{fig:Film}, in terms of pit depth versus time. It can be seen that, initially, the pit grows at approximately the same rate for all cases, independently of the stability of the film. This is because the film reduces corrosion kinetics progressively and pitting is initially occurring at high current densities (high $L_0$, diffusion-controlled corrosion). As the corrosion time increases, the cases with $k>0$ show how corrosion changes from diffusion-controlled to activation-controlled as $L$ decreases, up to the point where passivation prevents the material from corroding entirely. These results showcase the capabilities of the model in incorporating the role of film passivation in corrosion predictions.

\subsection{Hydrogen assisted cracking of a notched square plate}
\label{Sec:SENThydrogen}

\begin{figure}[H]
\centering
\noindent\makebox[\textwidth]{%
\includegraphics[scale=0.32]{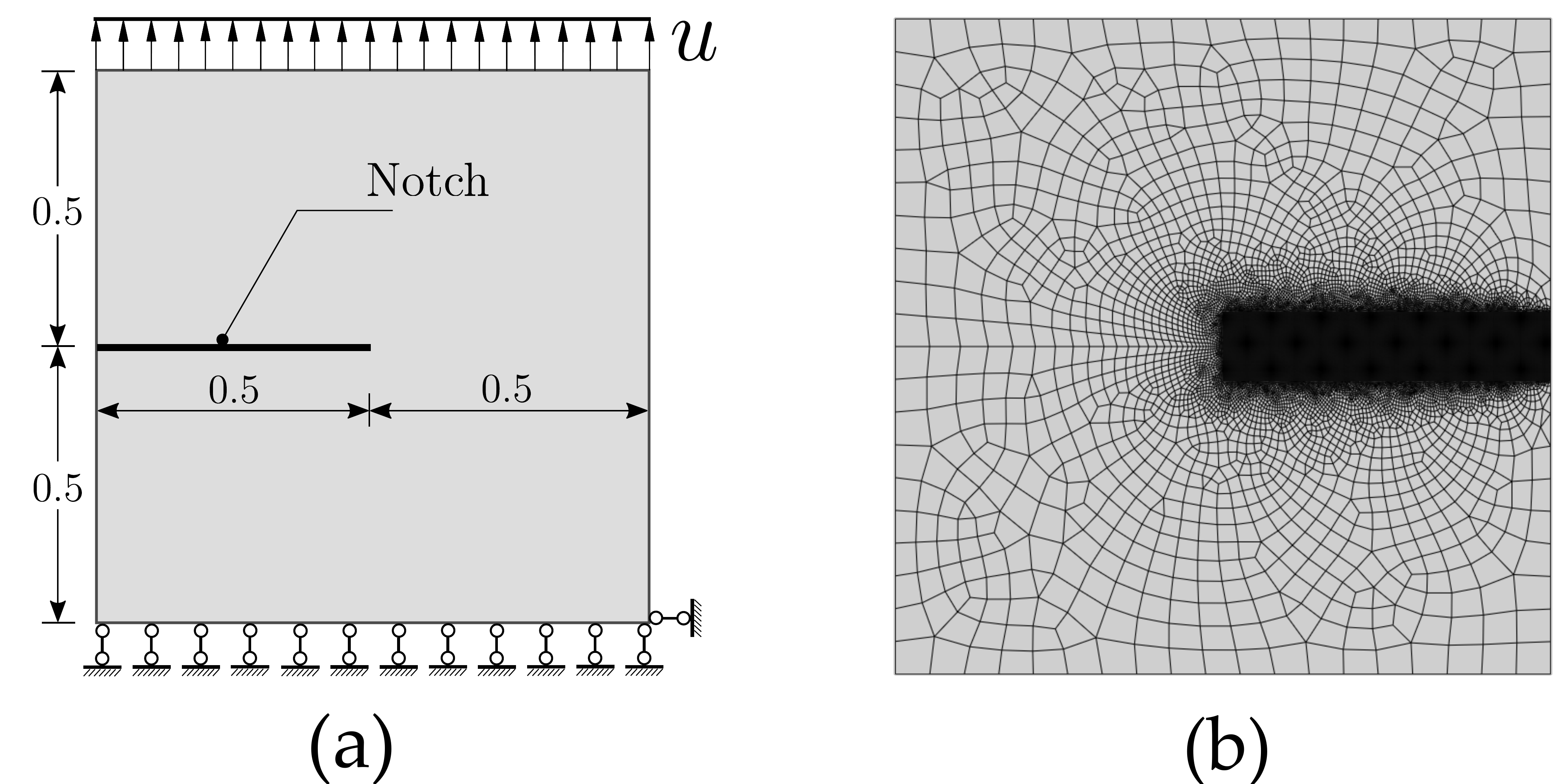}}
\caption{Hydrogen assisted cracking of a notched square plate: (a) geometry and mechanical loading, and (b) finite element mesh.}
\label{fig:cracked square}
\end{figure}

Insight and verification of the abilities of the model in predicting hydrogen embrittlement and cracking are attained by simulating the fracture of a notched square plate in a hydrogenous environment; a paradigmatic benchmark in the phase field fracture and phase field hydrogen embrittlement communities \citep{Miehe2010a,CMAME2018,Mandal2021}. As shown in Fig. \ref{fig:cracked square}a, we consider a 1 mm $\times$ 1 mm rectangular domain with a straight horizontal notch of length 0.5 mm at the mid-height of the left edge. The bottom edge is fixed and a vertical displacement is applied to the top edge. No corrosion effects are considered ($\phi_e=\phi_f$) and the relevant material properties are listed in Table \ref{tab:mecha_para}. Again, material parameters are chosen in agreement with previous computational studies \citep{Miehe2010a,CMAME2018}. The chemical initial and boundary conditions are as follows. First, the plate is uniformly charged with hydrogen such that $c_\mathrm{H} (t=0)= c_\text{env} \,  \forall \, \mathbf{x}$. Also, during the experiment, we enforce the penalty boundary condition such that $c_\mathrm{H}=c_\text{env}$ in the cracked regions. Selected values of $c_\text{env}$ are chosen to visualise the impact of increasing hydrogen content on fracture resistance; namely, 0 (no hydrogen), 0.1, 0.5 and 1 wt ppm. The mesh is refined in the expected crack propagation area, ensuring that the length scale parameter $\ell_f$ is at least 5 times larger than the characteristic element size (see Fig. \ref{fig:cracked square}b). Approximately 30,000 plane strain quadratic quadrilateral elements are used to discretise the plate. 

\begin{table} [!htbp]
  \centering
  \caption{Hydrogen assisted cracking of a notched square plate: material properties \citep{Miehe2010a,CMAME2018}.}
    \begin{tabular*}{\hsize}{@{}@{\extracolsep{\fill}}lll@{}}
    \toprule
    \textbf{Parameter} & \textbf{Value} & {\textbf{Unit}} \\
    \midrule
    Young’s modulus $E$ & 210,000  & MPa \\
    Poisson’s ratio $\nu$ & 0.3  & --- \\
    Critical energy release rate $G_c\left(0\right)$ & $2.7$ & $\text{kJ}/\text{m}^2$ \\
    Length scale parameter $\ell_f$ & $0.0075$ & mm \\
    Hydrogen diffusion coefficient $D_\mathrm{H}$ & $0.0127$ & $\mathrm{mm^2/s}$ \\
    Trap binding energy $\Delta g_b^0$ & $30$ & kJ/mol \\
    Partial molar volume $\overline{V}_\mathrm{H}$ & $2000$ & $\mathrm{mm^3/mol}$ \\
    Hydrogen damage coefficient $\chi$ & $0.89$ & --- \\
    \bottomrule
    \end{tabular*}
  \label{tab:mecha_para}
\end{table}

\begin{figure}[H]
\centering
\noindent\makebox[\textwidth]{%
\includegraphics[scale=0.42]{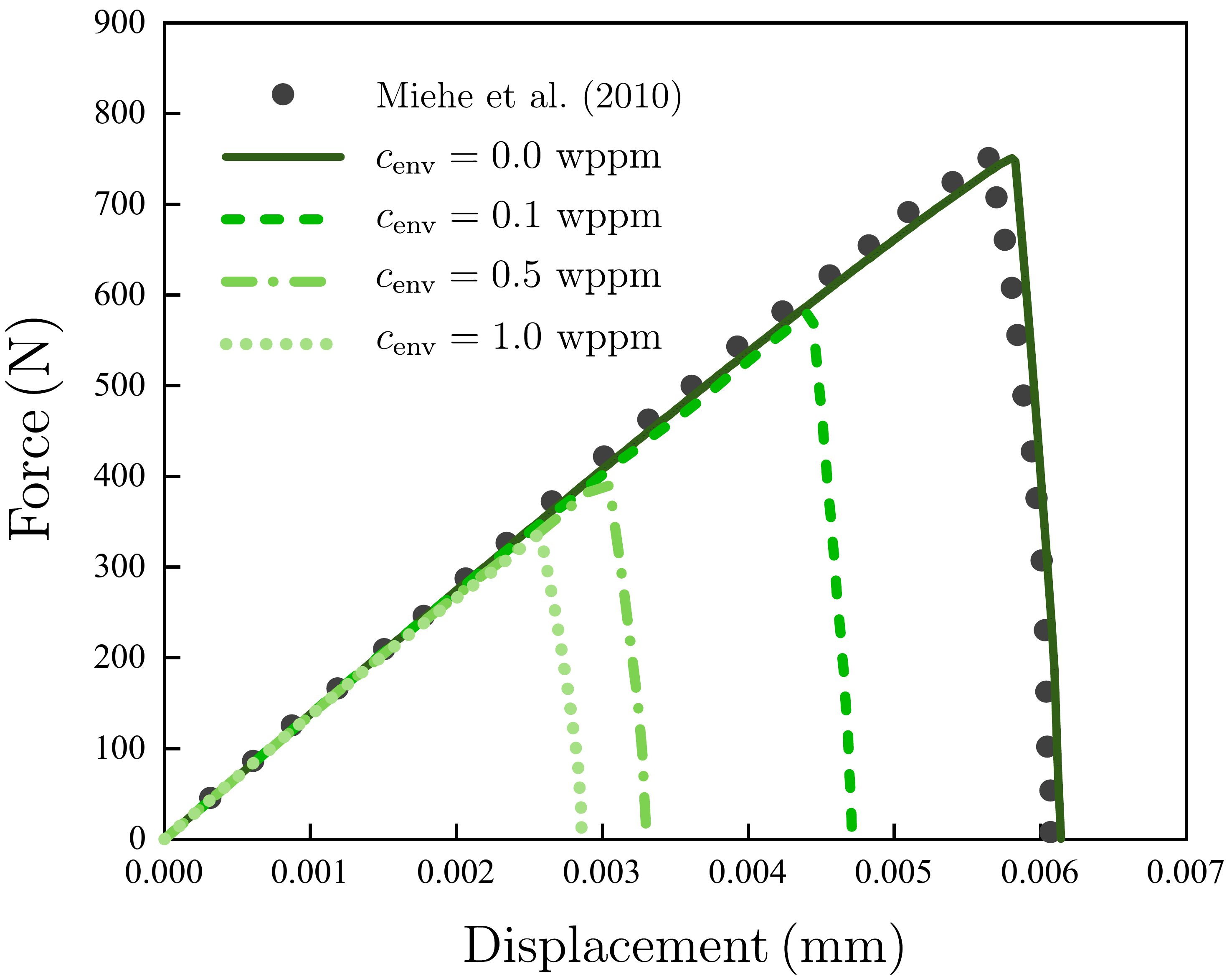}}
\caption{Hydrogen assisted cracking of a notched square plate: Force versus displacement curves for different hydrogen contents. The case without hydrogen ($c_\text{env}=0$) matches the results by \citet{Miehe2010a}.}
\label{fig:HE1}
\end{figure}

\begin{figure}[H]
\centering
\noindent\makebox[\textwidth]{%
\includegraphics[scale=0.28]{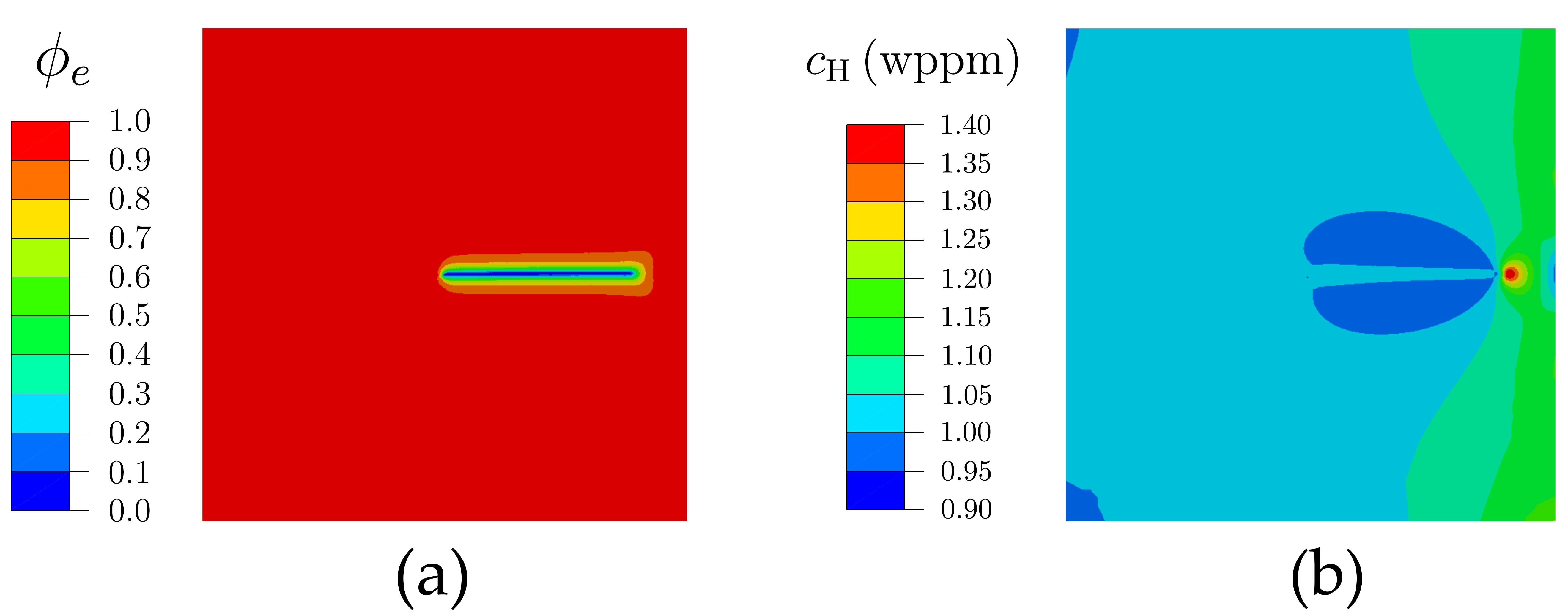}}
\caption{Hydrogen assisted cracking of a notched square plate: Contours of (a) damage variable $\phi_e$ and (b) hydrogen concentration $c_\text{H}$, when $c_\text{env}=1.0$ wppm and the remote displacement equals $u=0.0028$ mm.}
\label{fig:HE2}
\end{figure}

The force versus displacement curves obtained are shown in Fig. \ref{fig:HE1}. In the absence of hydrogen ($c_\text{env}=0$), predictions agree very well with the results by \citet{Miehe2010a}, validating the phase field fracture implementation. As the concentration of hydrogen increases, we observe a significant drop in the load carrying capacity of the plate, with crack growth initiating earlier and a notable reduction in the critical load. This is due to the hydrogen-induced degradation of the material toughness - see (T.4). Qualitatively, the results obtained agree with those reported in the literature. However, quantitative differences can be observed because, unlike in \citet{CMAME2018}, the damaged hydrostatic stress is used to drive hydrogen diffusion, see (T.5). Representative contours of the damage variable $\phi_e$ and hydrogen concentration $c_\text{H}$ are shown in Fig. \ref{fig:HE2}, for a given moment in time. In agreement with expectations, $c_\text{H}=c_\text{env}$ at the crack tip (as dictated by the \emph{moving} chemical boundary condition), and a higher hydrogen content is seen immediately ahead of the crack tip, where hydrostatic stresses are large. Thus, the present results validate the phase field fracture implementation and show that the interplay between hydrogen diffusion, fracture and mechanical deformation is adequately captured.

\subsection{Bending of stainless steel samples: transition from anodic dissolution to hydrogen embrittlement}
\label{Sec:bending test}

\begin{figure}[H]
\centering
\noindent\makebox[\textwidth]{%
\includegraphics[scale=0.75]{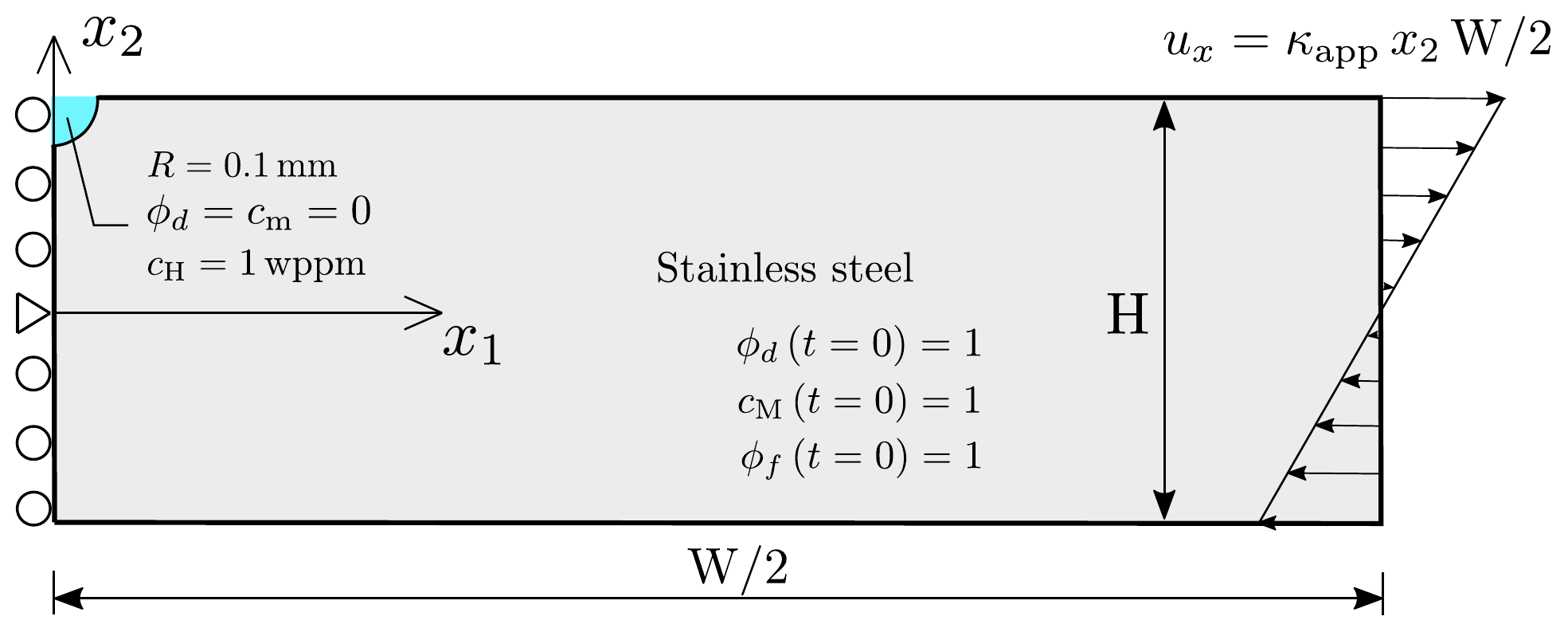}}
\caption{Bending of stainless steel samples: schematic with the configuration, dimensions and initial and boundary conditions.}
\label{fig:bending}
\end{figure}

We proceed to investigate the transition between anodic dissolution and hydrogen assisted damage mechanisms, as well as their interactions. For this purpose, we simulate the failure of a stainless steel beam of dimensions 10 mm $\times$ 2 mm undergoing pure bending while exposed to a corrosive environment. Taking advantage of symmetry, only half of the beam is modelled, as shown in Fig. \ref{fig:bending}, where the geometry, initial and boundary conditions are given. The beam height equals $\text{H}=2$ mm while the half-beam length is $\text{W}/2=5$ mm. The mechanical boundary condition corresponds to an applied curvature $\kappa_{\text{app}}$, such that the horizontal displacement at the beam ends evolves as,
\begin{equation}
    u_x = x_1 x_2 \kappa_{\text{app}} \,\,\,\,\,\,\,\, \text{at} \,\,\,\,\,\,\,\, x_1 \pm W/2
\end{equation}

\noindent Here, we choose to apply an initial curvature at $t=0$ and then increase its magnitude with time; namely: $\kappa_{\text{app}}=0.001 + 2\times10^{-8}\,t$. Similar to Section \ref{Sec:semicorro}, we assume that the stainless steel is well-protected from the corrosive environment except for a small semi-circular pit of radius $0.1$ mm, located at the centre of the top free surface and where the Dirichlet boundary conditions $\phi_d=0$, $c_\mathrm{M}=0$ and $c_\mathrm{H}=c_\mathrm{env}=1$ wppm are prescribed. The dissolution phase field length scale is chosen to be $\ell_d=0.06$ mm \citep{Mai2017}, while the mobility parameter equals $L_0=1\times10^{-5}$ mm$^2$/(N $\cdot$ s), to ensure activation-controlled corrosion conditions. The remaining electrochemical parameters are those reported in Table \ref{tab:corropara}. The hydrogen diffusion coefficient is defined as $D_\mathrm{H}=5.8\times10^{-6}$ mm$^2$/s, as reported in \citet{Luo2021}. In regard to the mechanical properties, the constitutive behaviour is characterised by a Young modulus of $E=190$ GPa, a Poisson’s ratio of $\nu=0.3$, a yield stress of $\sigma_y=400$ MPa, and a strain hardening exponent of $N=0.1$ \citep{Chen2006}. The fracture toughness is assumed to be $G_c\left(0\right)=1.2$ kJ/m$^2$. We adopt a material strength of $\sigma_c=800$ MPa which, according to (\ref{eq:sigma_c}), renders a phase field fracture length scale of $\ell_f=0.04$ mm. All the other parameters correspond to those listed in Table \ref{tab:mecha_para}. 

A passivation film deposits on the surface of the stainless steel sample and is ruptured if the local strain level is sufficiently high. This is captured by the film-rupture-dissolution-repassivation (FRDR) formulation presented in Section \ref{subsec:PhaseFieldCorr}. The interplay between film rupture and the environment is characterised by the parameters $k=0.0001$, $t_0=50$ s, and $\varepsilon_f=0.003$. Moreover, it is well-known that the presence of a passive film can prevent the uptake of hydrogen (see, e.g., \citealp{Schomberg1996}). We incorporate this film-hydrogen interaction by defining, when the passive film begins to stabilise ($t_i>t_0$), a sufficiently low hydrogen diffusion coefficient $D_\mathrm{H}$ along the electrolyte-metal interface ($\phi_d\leqslant0.95$); i.e.,
\begin{equation} \label{eq:FilmHinter}
    \left\{
\begin{aligned}
    & D_\mathrm{H} \to 0, \,\,\,\,\,\,\,\,\,\,\,\,\text{if} \,\,\, t_i > t_0 \,\,\,\text{and}\,\,\,\phi_d\leqslant0.95 \\
    & D_\mathrm{H} \equiv D_\mathrm{H} , \,\,\,\,\,\,\,\,\,\text{otherwise} \\
\end{aligned}
\right.
\end{equation}

\noindent For numerical reasons, $D_\mathrm{H}$ is not chosen to be exactly zero but equal to a small value. Specifically, a magnitude of $D_\mathrm{H}=1 \times 10^{-10}$ mm$^2$/s achieves the desired effect. One should note that the uptake of hydrogen is only hindered at the interface; the formulation still captures the condition $c_\mathrm{H}\equiv c_\mathrm{env}$ in all regions of the evolving electrolyte due to the penalty boundary condition (\ref{r_H}). A total of approximately 13,000 quadrilateral quadratic elements with reduced integration are used to discretise the model, with the mesh being particularly refined in the expected SCC region.\\

\begin{figure}[H]
\centering
\noindent\makebox[\textwidth]{%
\includegraphics[scale=0.2]{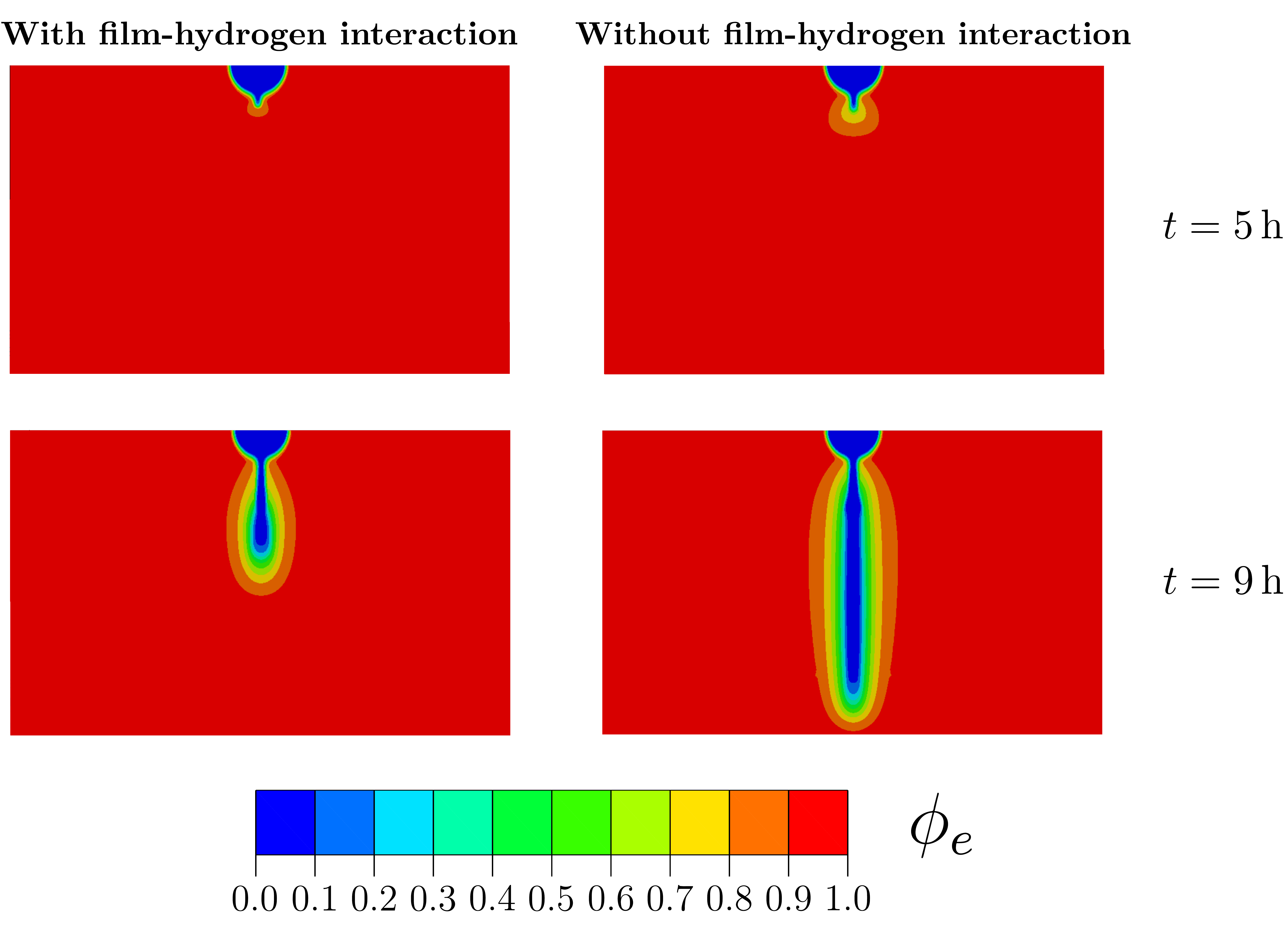}}
\caption{Bending of stainless steel samples: predictions of the evolution of the SCC damage region considering and neglecting the interaction between hydrogen uptake and film passivation.}
\label{fig:phie_compare}
\end{figure}

Numerical experiments are conducted to investigate the influence of the various parameters at play. First, we compare the predictions obtained with and without the film-hydrogen interaction defined in (\ref{eq:FilmHinter}). The results are shown in Fig. \ref{fig:phie_compare} for two instants of time, $t=5$ and $t=9$ hours. It can be observed that for short time scales ($t\leqslant 5$ h), predictions are relatively similar, suggesting that SCC growth is dominated by anodic dissolution. However, when there is enough time for hydrogen to ingress and diffuse in the sample, then differences become significant. The length of the SCC crack is twice as large if the model does not account for the interplay between hydrogen uptake and film passivation. 

\begin{figure}[H]
\centering
\noindent\makebox[\textwidth]{%
\includegraphics[scale=0.22]{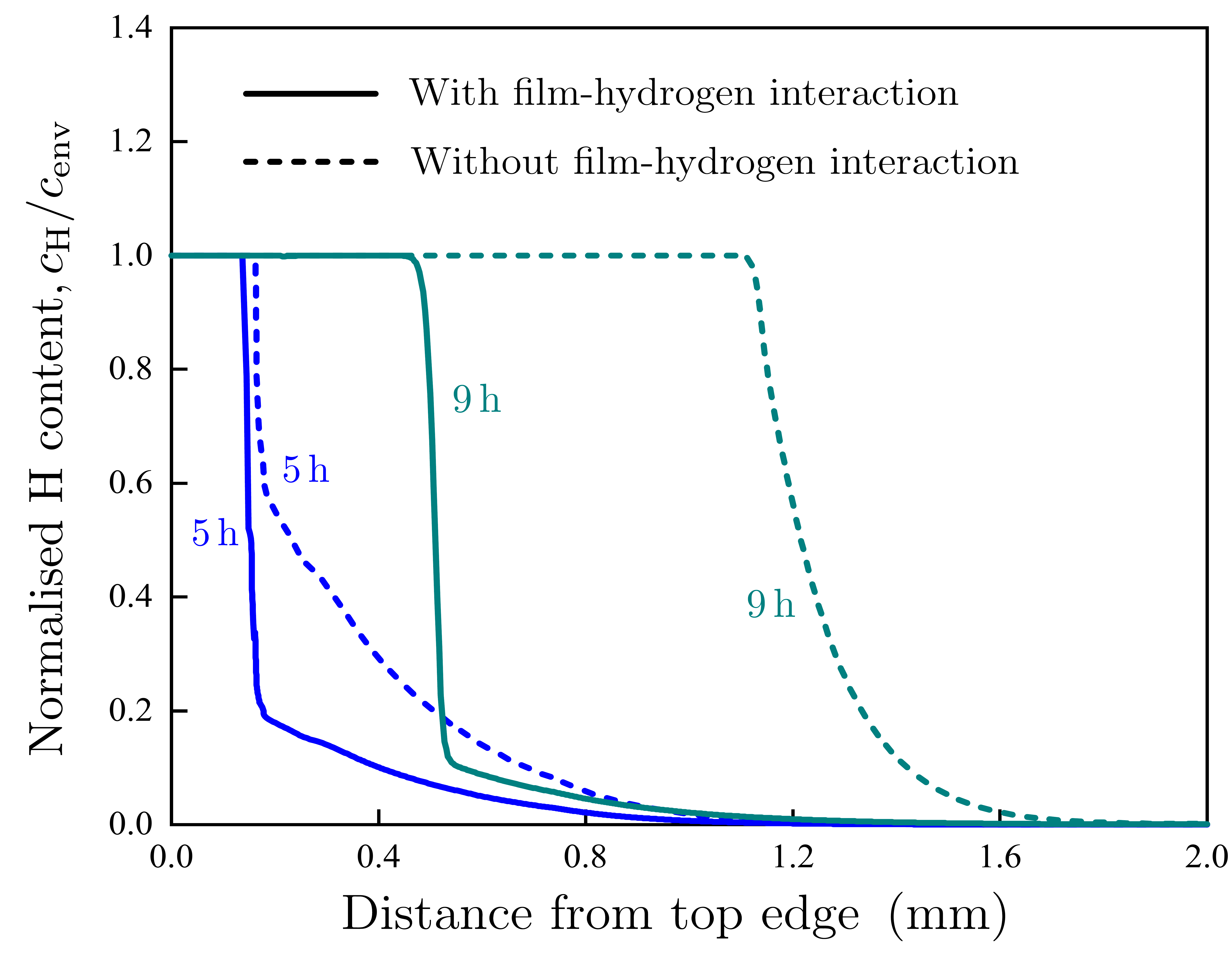}}
\caption{Bending of stainless steel samples: normalised hydrogen content along the extended SCC propagation path as predicted considering and neglecting the interaction between hydrogen uptake and film passivation.}
\label{fig:H_compare}
\end{figure}

Further insight into this film-hydrogen uptake interplay is gained by plotting the normalised distribution of hydrogen content, see Fig. \ref{fig:H_compare}. The SCC defect length at each reported time period can be readily identified from the hydrogen distribution plots, as it is associated with a drop in $c_\mathrm{H}$ from $c_\mathrm{env}$. Thus, it can be seen that the SCC defect lengths predicted by models with and without the film-hydrogen interaction condition (\ref{eq:FilmHinter}) are very close for $t=5$ h, but significant differences are observed after 9 h. This is due to the shift in damage driving force mechanism. For $t\leqslant$5 h, the growth of the SCC defect is driven by anodic dissolution. However, as time increases, two effects become relevant. First, the SCC defect sharpens due to the role of the mechanics-enhanced corrosion terms (\ref{eq:L_cycle}) and this is further enhanced as a result of the localised rupture of the passive film. Second, the applied load increases with time. These two features lead to a localised stress raise, which increases the hydrogen content (\ref{eq:flux_H}) and the energy release rate. This triggers a shift between anodic dissolution- and hydrogen embrittlement-driven SCC defect growth. Additionally, the results in Fig. \ref{fig:H_compare} show how the condition (\ref{eq:FilmHinter}) influences hydrogen uptake, with a sharp drop in the hydrogen content being observed for the cases in which the passivation layer impedes hydrogen ingress. However, one should note that hydrogen ingress still takes place to a certain extent, as there is a time $t_0$ in each repassivation-rupture cycle where the film has not had enough time to stabilise and protect. 

\begin{figure}[H]
\centering
\noindent\makebox[\textwidth]{%
\includegraphics[scale=0.22]{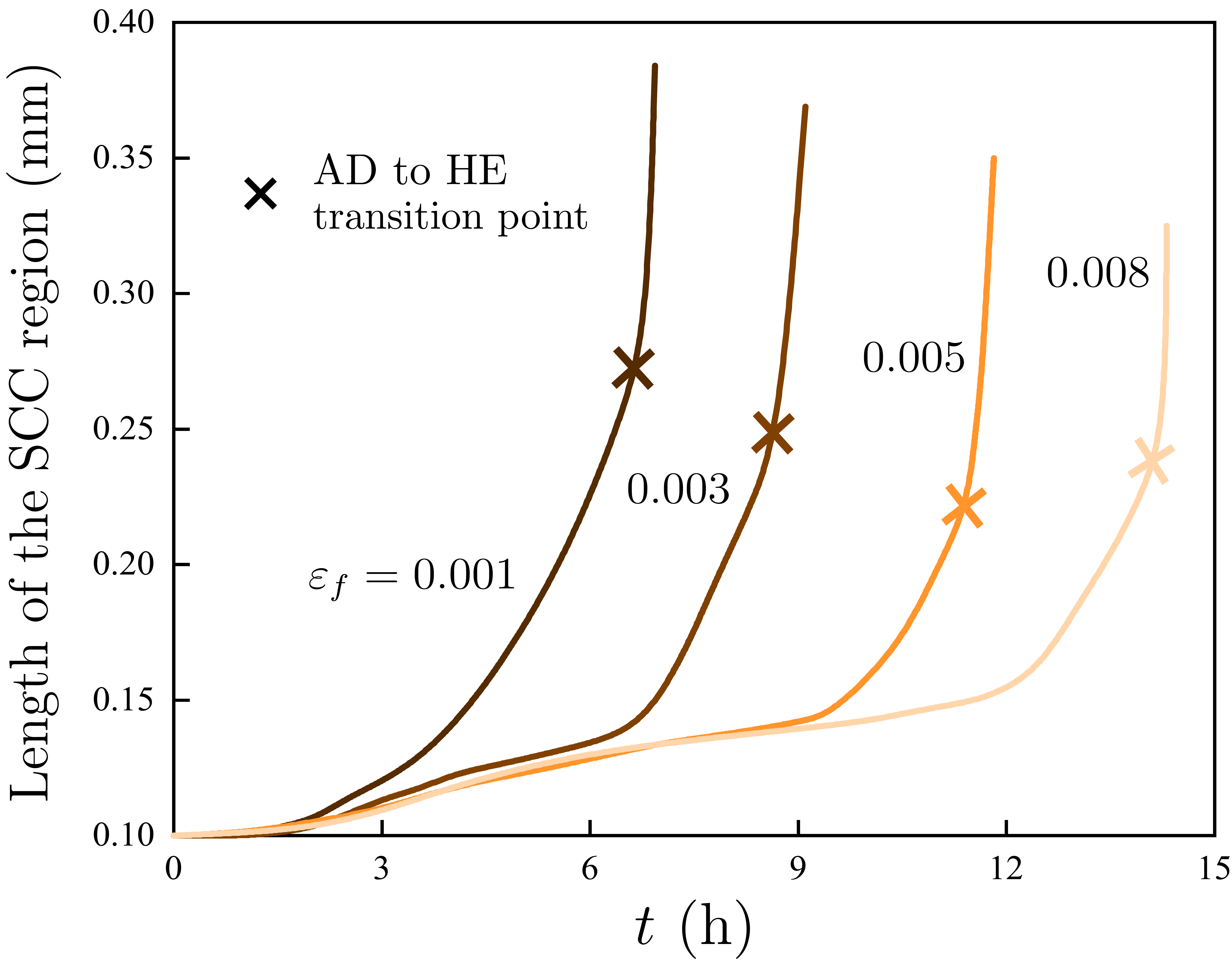}}
\caption{Bending of stainless steel samples: Length of the SCC region as a function of time for four representative failure strain values. A cross is used to denote the time at which the growth of the SCC defect transitions from being driven by anodic dissolution to being dominated by hydrogen embrittlement mechanisms ($\phi_f\leqslant0.05$).}
\label{fig:SCCtrans}
\end{figure}

We conclude this case study by investigating how the film rupture resistance influences the transition from anodic dissolution to hydrogen embritlement SCC defect growth. For this purpose, it is assumed that SCC defect growth is dominated by hydrogen embrittlement mechanisms if $\phi_f\leqslant0.05$. The results obtained are reported in Fig. \ref{fig:SCCtrans} in terms of the length of the SCC region versus time, with a cross being used to highlight the transition point from anodic dissolution dominance to hydrogen embrittlement dominance. Four curves are shown, each for a different film failure strain $\varepsilon_f$. In all cases, the role of the passivation film in preventing hydrogen uptake is accounted for; as elaborated above, this hinders hydrogen ingress but does not completely preclude it, due to the time required for the film to stabilise. In agreement with expectations, the results reveal that the SCC defect grows faster if the fracture resistance of the film is smaller. The rate of SCC defect growth increases with time due to the sharpening of the SCC pit - the tip of the defect acts as a stress concentrator, which favours a very localised rupture of the film and increases dissolution rates in regions of high stresses and strains. Defect growth rates are further augmented when hydrogen mechanisms dominate stress corrosion cracking, with the crack eventually propagating in an unstable manner. The interplay between $\varepsilon_f$ and the transition to hydrogen embrittlement is more complex. The sensitivity of the transition time to $\varepsilon_f$ agrees with expectations; hydrogen embrittlement becomes dominant later in time for larger values of $\varepsilon_f$, as these lead to smaller hydrogen exposure times and lower SCC growth rates overall. However, $\varepsilon_f$ has a dual influence on the length of the SCC defect at which hydrogen embrittlement dominance is attained. On the one hand, longer times lead to higher values of the energy release rate, favouring hydrogen-assisted mechanical cracking relative to dissolution-driven defect growth. On the other hand, the higher $\varepsilon_f$ values associated with larger failure times result in a reduced environmental exposure and thus a smaller size of the hydrogen-degraded region. 

\subsection{SCC threshold predictions in artificial and biologically-active seawater}
\label{Sec:pit-to-crack transition}

\begin{figure}[H]
\centering
\noindent\makebox[\textwidth]{%
\includegraphics[scale=0.17]{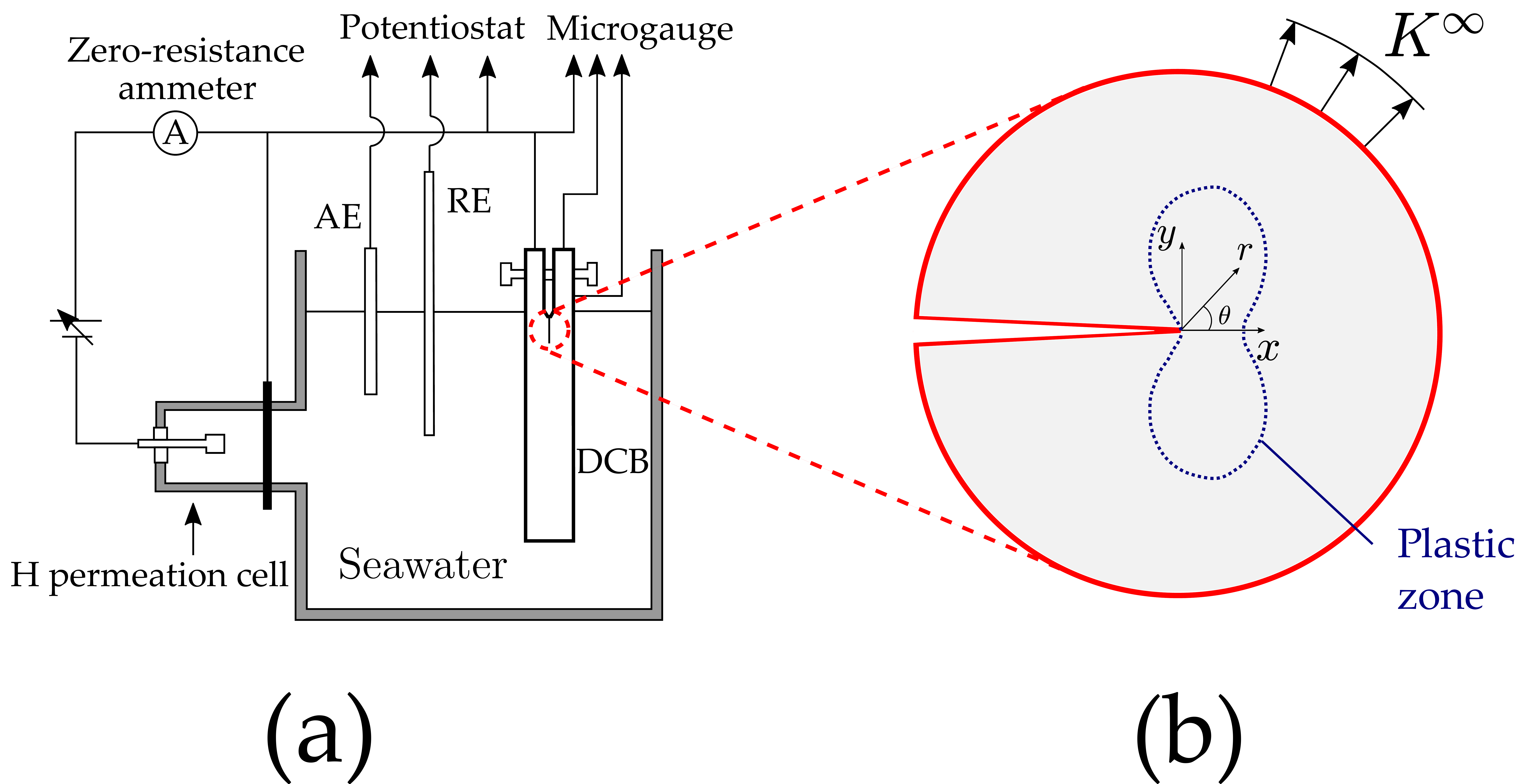}}
\caption{SCC threshold predictions as a function of the environment: Schematic illustration of (a) the DCB test configuration \citep{Robinson1994}, and (b) the boundary layer model to reproduce crack tip loading conditions under the assumption of small scale yielding.}
\label{fig:DCB}
\end{figure}

The experiments by \citet{Robinson1994} on heat-treated steel 690 are simulated and extended to illustrate the ability of the model to predict cracking thresholds under the dominance of anodic dissolution or hydrogen embrittlement mechanisms, as well as in the transition from one to the other. Steel 690 samples were exposed to sterile artificial seawater (SAS) and biologically active seawater (BAS); the existence of bacteria in BAS increases sulfide concentration and consequently promotes hydrogen uptake. Experiments were conducted at a wide range of cathodic protection potentials and stress intensity thresholds $K_{th}$ were measured in each scenario. As sketched in Fig. \ref{fig:DCB}a, the experiments were conducted using a double-cantilever beam (DCB) configuration with an initial crack. Given that small scale yielding conditions prevail, we simplify the loading configuration by using a so-called boundary layer model, where a remote $K_I$-field is prescribed, characterising the stress state near the crack in the DCB test (see Fig. \ref{fig:DCB}b). Thus, the loading conditions are defined by simulating a semi-circular geometry and prescribing the displacements on the outer surface in agreement with \citet{Williams1957} elastic solution; i.e., the horizontal and vertical displacements of the outer nodes respectively read,
\begin{equation}
    u_x = K_I \frac{1+\nu}{E} \sqrt{\frac{r}{2 \pi}} \left( 3- 4 \nu - \cos \theta \right) \cos \left( \frac{\theta}{2} \right)
\end{equation}
\begin{equation}
    u_y = K_I \frac{1+\nu}{E} \sqrt{\frac{r}{2 \pi}} \left( 3- 4 \nu - \cos \theta \right) \sin \left( \frac{\theta}{2} \right)
\end{equation}

\noindent where $r$ and $\theta$ are the coordinates of a polar coordinate system centred at the crack tip. 

The electrochemical parameters listed in Table \ref{tab:corropara} are adopted unless otherwise stated. A phase field corrosion length scale of $\ell_d=0.05$ mm is used. According to \citet{Tian2018}, no passivation occurs in low-alloy steels across a wide range of artificial seawater environments, such that $k=0$ in (\ref{eq:L_cycle}) is considered throughout this analysis. The constitutive parameters for steel 690 are determined as $E=210$ GPa, $\nu=0.3$, and $N=0.1$ based on the uniaxial stress-strain data reported by \cite{Billingham2003}, while the yield strength is taken to be $\sigma_\text{y}=1032$ MPa, as reported by \citet{Robinson1994}. In the absence of hydrogen, we consider a material toughness of $G_c\left(0\right)=10$ kJ/m$^2$, which is estimated from the stress intensity threshold measured in an inert environment \citep{Robinson1994}. The phase field length scale can be chosen in agreement with (\ref{eq:sigma_c}), for known values of $E$, $G_c$ and the material strength $\sigma_c$. No data for $\sigma_c$ is available but given the low hardening of these alloys, the ultimate tensile strength is expected to be between $\sigma_y$ and 1.5 times $\sigma_y$. These would give $\ell_f$ values of 0.09 and 0.21 mm; accordingly we choose the intermediate value $\ell_f=0.15$ mm. Mimicking the experiments, a very slow load rate is considered, $\dot{K}_I=10^{-5}$ $\mathrm{MPa\sqrt{m}/s}$, such that there is sufficient time for the hydrogen to achieve steady state conditions. \citet{Robinson1994} report $K_{th}$ versus hydrogen content curves, as the focus was on cathodic protection conditions, where hydrogen embrittlement is dominant. The cathodic potential is applied from $-0.85$ V to $-1.25$ V with an interval of $0.1$ V, such that a total of ten stress intensity thresholds are recorded with different hydrogen concentrations. We will extend the range of applied potentials beyond the experiments to predict the transition to dissolution-driven SCC, but begin by reproducing the testing data for hydrogen embrittlement conditions. For the low applied potentials of the experimental campaign, material dissolution is likely to be negligible and accordingly a very small interface kinetics coefficient of $L_0=10^{-15}$ $\mathrm{mm^2/(N \cdot s)}$ is considered, with $\phi_d=c_\text{M}=0$ defined as initial conditions at the crack tip node. The hydrogen initial and boundary conditions involve the definition of an initial uniform hydrogen concentration throughout the sample $c_\mathrm{H} (t=0)= c_\text{env} \,  \forall \, \mathbf{x}$, and the application of the boundary condition $c_\mathrm{H} = c_\text{env}$ in the crack surfaces. A diffusion coefficient of $D_\mathrm{H}=2\times10^{-6}$ $\mathrm{mm}^2/\mathrm{s}$ is defined, as reported by \cite{Robinson1994}. The mesh in the vicinity of the crack is refined to be at least 10 times smaller than both the interface thickness $\ell_d$ and the length scale parameter $\ell_f$, so as to ensure the mesh insensitive results. Only half of the boundary layer model is considered, taking advantage of symmetry, and the model is discretised with approximately 65,000 quadratic quadrilateral elements with reduced integration.\\

\begin{figure}[H]
\centering
\noindent\makebox[\textwidth]{%
\includegraphics[scale=0.43]{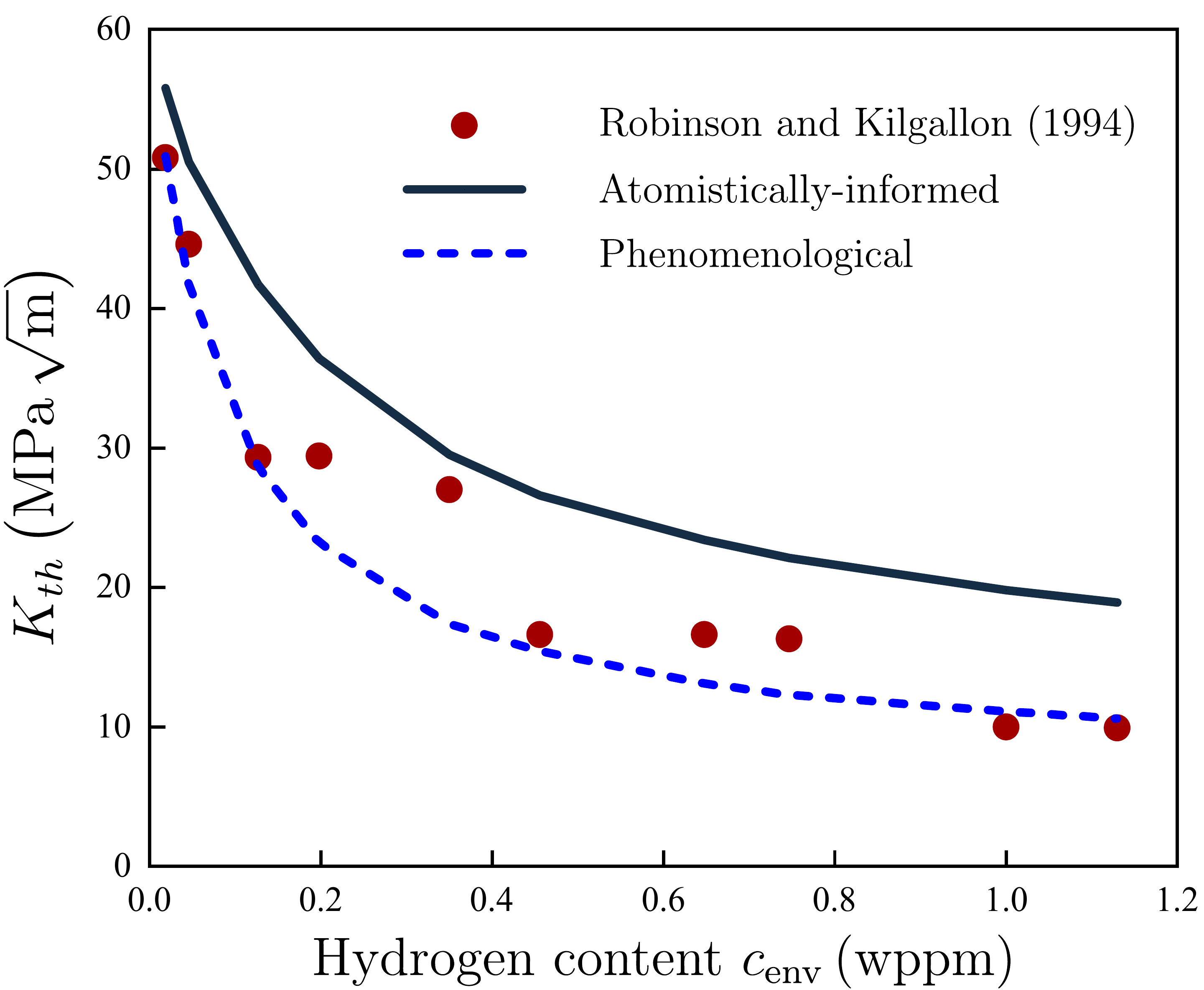}}
\caption{SCC threshold predictions as a function of the environment: experimental \citep{Robinson1994} and modelling predictions of $K_{th}$ versus hydrogen concentration $c_{\text{env}}$. The numerical results have been obtained with an atomistically-inspired hydrogen degradation law, Eq. (\ref{eq:G_c}), and with a phenomenological one, Eq. (\ref{eq:G_c2}).}
\label{fig:Kth_cH2}
\end{figure}

Experimental and computational predictions of stress intensity threshold $K_{th}$ versus hydrogen content are shown in Fig. \ref{fig:Kth_cH2}. The numerical results have been obtained considering two hydrogen degradation laws: (i) the atomistically-informed one presented in Section \ref{Sec:FractureInterfaceEnergyDensity}, with $\chi=0.98$, and (ii) a phenomenological degradation law that provides a good fit to the experiments. The latter is given by the following higher order function:
\begin{equation}\label{eq:G_c2}
    \begin{aligned}
      G_c\left(\theta\right)=\left( 0.0285\theta^4-0.3532\theta^3+1.433\theta^2-2.0934\theta+1\right)\, G_c\left(0\right)
    \end{aligned}
\end{equation}

\begin{figure}[H]
\centering
\noindent\makebox[\textwidth]{%
\includegraphics[scale=0.45]{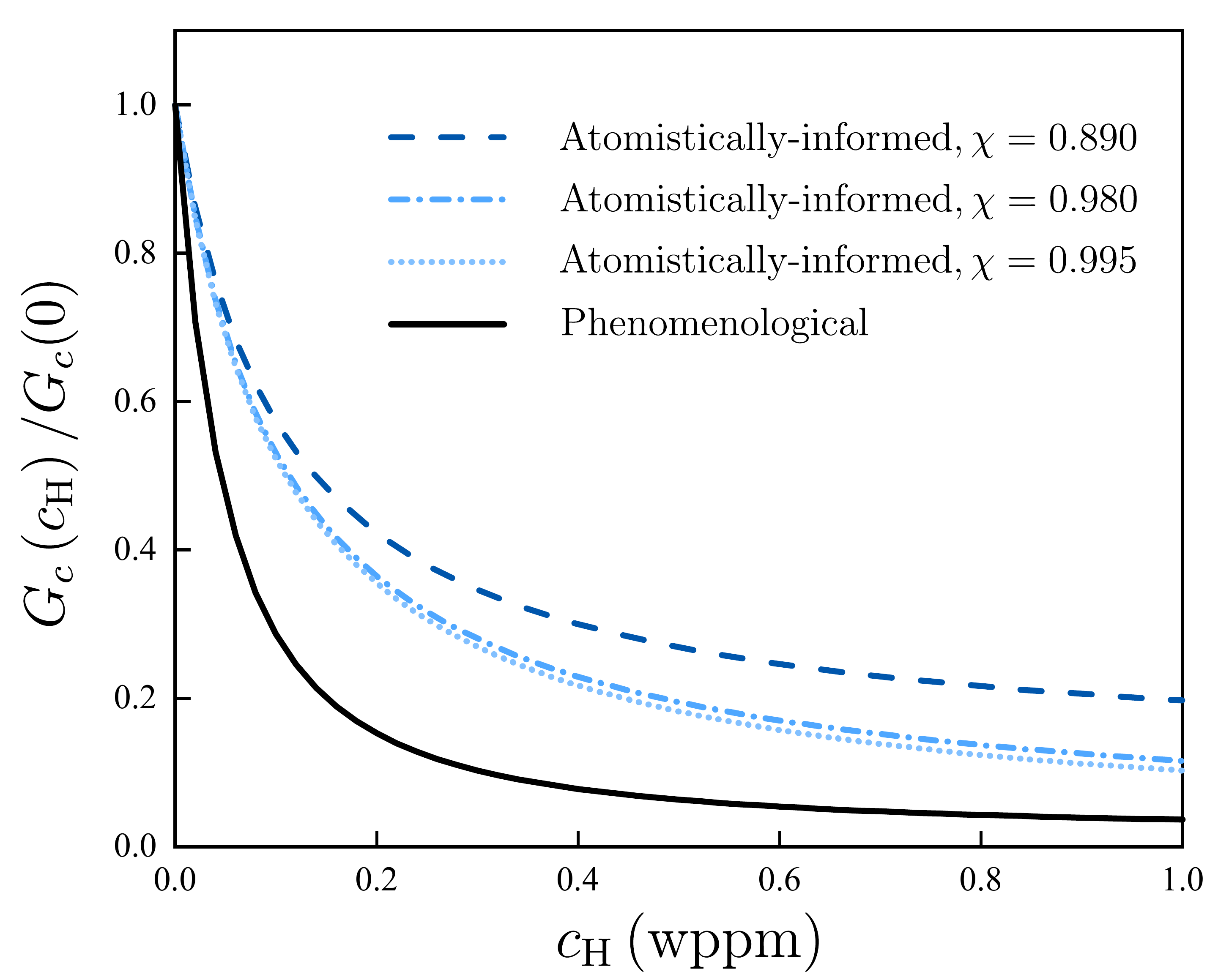}}
\caption{SCC threshold predictions as a function of the environment: Sensitivity of the material toughness to the hydrogen content, as predicted by both the atomistically-informed degradation law (\ref{eq:G_c}) and by the phenomenological relation (\ref{eq:G_c2}). The atomistic approach uses an interface binding energy of 30 kJ/mol.}
\label{fig:degradation2}
\end{figure}

Both degradation laws lead to the same qualitative outcome: a significant reduction in $K_{th}$ with increasing hydrogen content. However, the atomistically-informed one slightly underpredicts the degree of embrittlement, independently of the choice of $\chi$ (the hydrogen damage coefficient). The sensitivity of the material toughness as a function of the hydrogen concentration is shown in Fig. \ref{fig:degradation2}. It can be seen that the atomistic relation saturates as $\chi \to 1$ at higher levels of $G_c$, relative to the degradation law derived from experiments. However, the first principles predictions are very sensitive to the choice of the interface binding energy - see Eq. (\ref{eq:theta}). Here, a value of 30 kJ/mol has been chosen based on an average of the experimental data available for grain boundaries but measurements can range from 17 to 59 kJ/mol \citep{Asaoka1977,Choo1982}. Reducing the uncertainty in the characterisation of trap binding energies is arguably the bottleneck for parameter-free mechanistic modelling of hydrogen assisted cracking. 

Now, let us model a wider range of environments, as characterised by the applied potential $E_p$, to additionally consider conditions of free corrosion and anodic-driven SCC. The polarisation curves reported by \citet{Tian2018} for similar conditions (material, solutions) will be used to determine the free corrosion potential $E_{\text{corr}}$, and a higher $E_p$, such that anodic dissolution becomes the dominant mechanism for cracking. To capture the sensitivity to the applied potential, we will exploit the relation between the interface mobility parameter $L$ and the corrosion current density $i$, as elaborated next. Let us assume a stress-free condition, leaving aside the dependency of the mobility coefficient on mechanical fields (i.e., $L \equiv L_0$). Also, let us assume activation-controlled corrosion conditions (i.e., a sufficiently small $L \equiv L_0$). Under activation-controlled conditions, the corrosion velocity $v_n$ is related to the current density $i$ through Faraday's second law for electrolysis:
\begin{equation}\label{eq:Faraday}
    v_n = \frac{i}{z F c_\mathrm{solid}}
\end{equation}

\noindent where $z=2.1$ is the average charge number and $F$ is Faraday's constant. Recall now the Allen-Cahn type equation used to evolve the corrosion phase field, (T.2). Since, in activation-controlled corrosion, $v_n \equiv \text{d}\phi_d/\text{d}t$, the combination of (\ref{eq:Faraday}) and (T.2) implies that the interface mobility coefficient is proportional to the current density: $L \propto i$. It follows immediately that for any corrosion current density (e.g., $i_1$), the associated mobility coefficient ($L_1$) can be readily determined, provided that the proportionality constant is known for another current (e.g., $L_2/i_2$); i.e.,
\begin{equation}\label{eq:Liratio}
   L_1 = i_1\frac{L_2}{i_2}
\end{equation}

Accordingly, one can run a uniform corrosion simulation with any sufficiently small $L$ value that leads to activation-controlled corrosion conditions, compute the velocity of the corrosion front $v_n$ and use (\ref{eq:Faraday}) to estimate the current density associated with that $L$ \citep{Mai2016}. Once the $L/i$ ratio is known, Eq. (\ref{eq:Liratio}) can be used to obtain the interface mobility coefficient for any given corrosion current density. Accordingly, we use the polarisation data by \citet{Tian2018} and follow this strategy to determine the mobility coefficient associated with the free corrosion current density $i_{corr}$ for two environments, R2 and AS2, that respectively resemble the sterile artificial seawater and biologically active seawater solutions. First, a simulation with a small value of $L=L_0=2\times10^{-8}$ $\mathrm{mm^2/(N \cdot s)}$ is conducted, leading to $v_n=3.3\times10^{-9}$ $\mathrm{mm/s}$ and, \textit{via} (\ref{eq:Faraday}), $i=9.6\times10^{-8}$ $\mathrm{A/mm^2}$ ($L/i=0.2083$). Now, we use (\ref{eq:Liratio}) to find a value of $L_{corr}=L_0=3.3\times10^{-10}$ $\mathrm{mm^2/(N \cdot s)}$, as the mobility coefficient associated with the free corrosion current density $i_{corr}$ for sterile artificial seawater (where $i_{corr}=1.59\times10^{-9}$ $\mathrm{A/mm^2}$), and of $L_{corr}=L_0=3.33\times10^{-9}$ $\mathrm{mm^2/(N \cdot s)}$ for biologically active seawater (where $i_{corr}=1.6\times10^{-8}$ $\mathrm{A/mm^2}$). These values correspond to the corrosion potential, which equals $E_{corr}=-0.74$ V$_{SCE}$ for sterile artificial seawater and $E_{corr}=-0.65$ V$_{SCE}$ for biologically active seawater. A similar but slightly different strategy is used to determine the mobility coefficients associated with the anodic condition. Here, we utilise the corrosion velocity measurements by \citet{Tian2018}, use that value to compute $i$ \textit{via} (\ref{eq:Faraday}) and then determine $E_p$ using their polarisation curves and the associated mobility coefficient through (\ref{eq:Liratio}). This leads to values of $E_p=-0.68$ V$_{SCE}$, $L_0=9.2\times10^{-8}$ $\mathrm{mm^2/(N \cdot s)}$ and $E_p=-0.63$ V$_{SCE}$, $L_0=3.1\times10^{-7}$ $\mathrm{mm^2/(N \cdot s)}$ for sterile artificial seawater and biologically-active seawater, respectively. To estimate the hydrogen content across all applied potentials we apply a linear fit to the data provided by \citet{Robinson1994} and extrapolate.

\begin{figure}[H]
\centering
\noindent\makebox[\textwidth]{%
\includegraphics[scale=0.45]{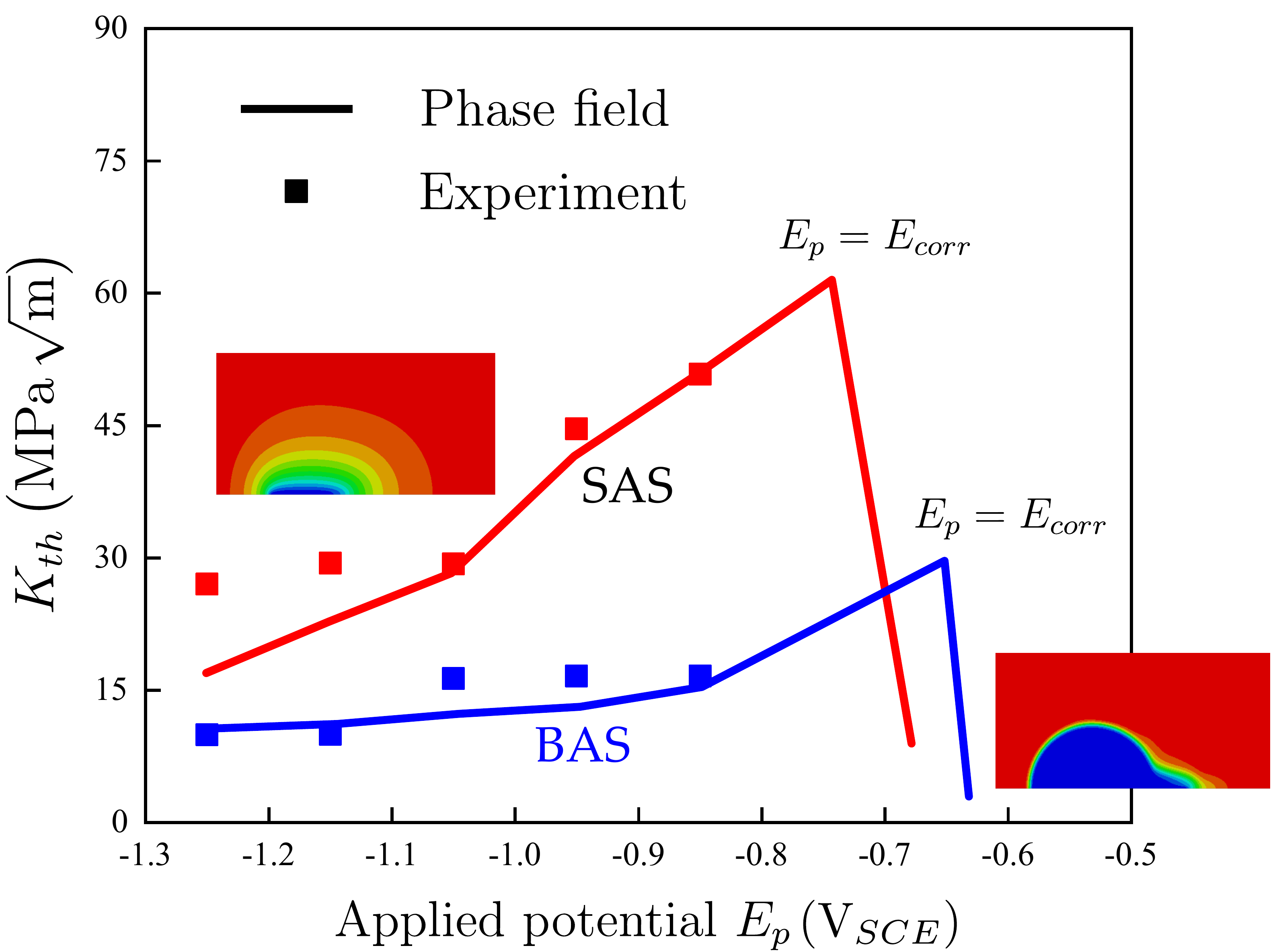}}
\caption{SCC threshold predictions as a function of the environment: Stress intensity threshold $K_{th}$ versus applied potential $E_p$, for both sterile artificial seawater (SAS) and biologically active seawater (BAS). The cathodic region includes experimental data by \citet{Robinson1994}.}
\label{fig:Kth_V}
\end{figure}

The results obtained are shown in Fig. \ref{fig:Kth_V}, in terms of cracking thresholds $K_{th}$ versus applied potential $E_p$, for both sterile artificial seawater (SAS) and biologically active seawater (BAS). The experimental data obtained by \citet{Robinson1994} for the cathodic region is also shown. The model agrees quantitatively with experiments in the cathodic region and qualitatively captures the expected trends for $E_{corr}$ and higher applied potentials. Namely, $K_{th}$ increases with applied potential until reaching the peak at $E_p=E_{corr}$, and then drops again as the applied potential enters the regime where cracking is driven by anodic SCC mechanisms \citep{Lee2007,Pioszak2017}. Thus, the model can satisfactorily predict the sensitivity of cracking thresholds to the environment across all regimes: cathodic, corrosion and anodic potentials.

\subsection{SCC tube testing of Al alloys \citep{Gruhl1984}: hydrogen embrittlement vs anodic dissolution}
\label{Sec:GruhlExpt}

\begin{figure}[H]
\centering
\noindent\makebox[\textwidth]{%
\includegraphics[scale=0.46]{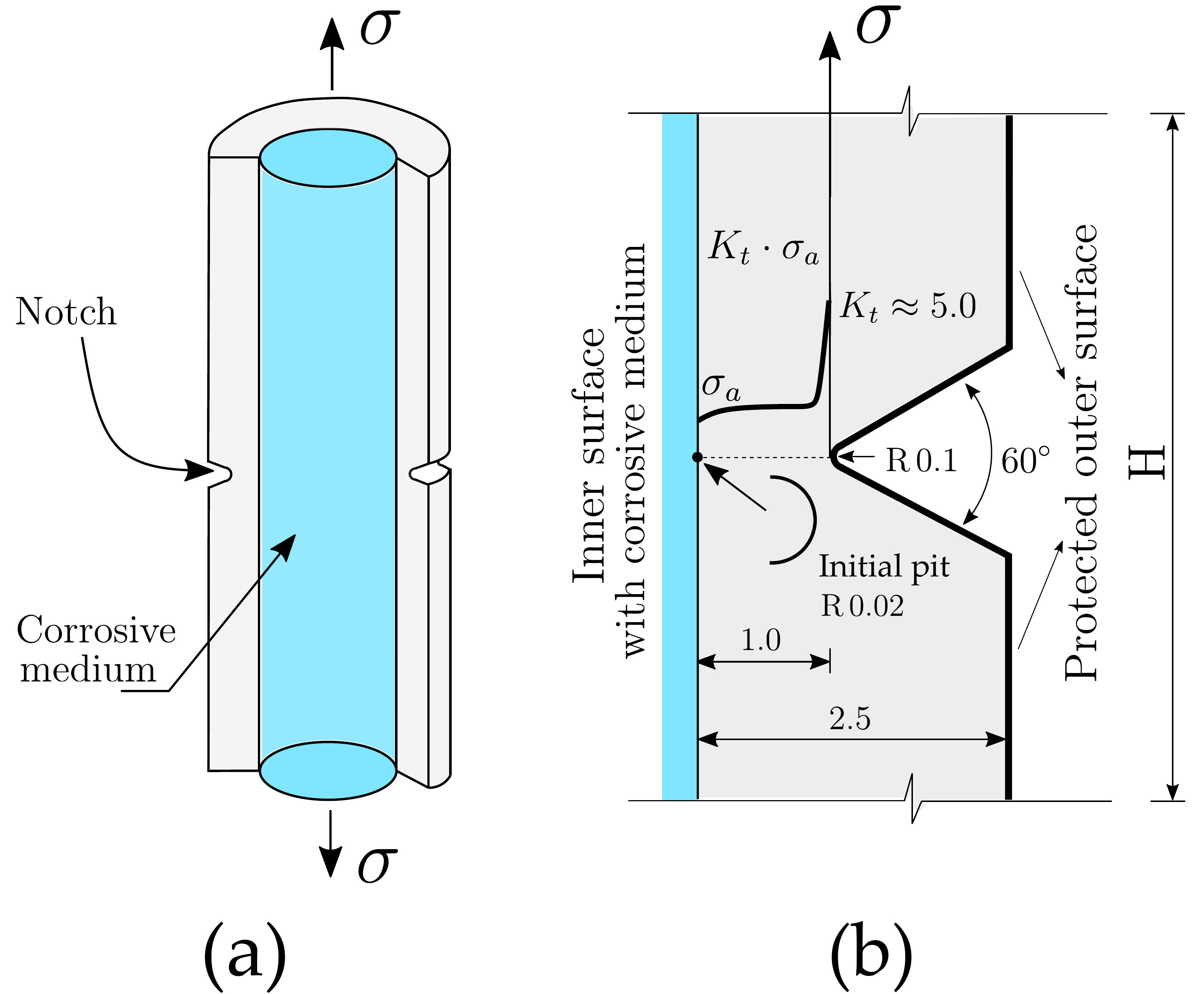}}
\caption{\citet{Gruhl1984} experiments: (a) Testing configuration, and (b) detailed view of the geometry, with dimensions given in mm. Anodic-driven damage mechanisms should trigger cracks in the surface in contact with the environment, while cracking due to hydrogen embrittlement will be favoured near the notch region, where hydrogen accumulates due to higher hydrostatic stresses.}
\label{fig:Expmodel2}
\end{figure}

Finally, we gain insight into the competition between anodic dissolution and hydrogen damage mechanisms by modelling the classic \citet{Gruhl1984} experiments on Al alloys. As shown in Fig. \ref{fig:Expmodel2}, \citet{Gruhl1984} shed light into the nature of SCC in Al alloys by conducting tensile tests on cylindrical samples that were exposed to a corrosive environment on their inner surface and had a notch in the outer one. If stress corrosion cracking was driven by anodic mechanisms, then cracks should initiate from the inner surface. However, if SCC failures were governed by hydrogen embrittlement, then cracking should initiate close to the notch, in a region far from the corrosive medium but where hydrogen could diffuse and accumulate due to the role that the notch plays as a stress concentrator. 

More specifically, \citet{Gruhl1984} conducted experiments on AlZn5Mg3 using an aqueous solution containing 2\% NaCl and 0.5\% Na2CrO4, with the pH value being adjusted to equal 3 (i.e., corrosion regime). The samples were hollow cylinders with the dimensions shown in Fig. \ref{fig:Expmodel2}, and had outer notches of 1.5 mm depth and 0.1 mm tip radius. Only the inner surface of the samples was exposed to the environment, while the outer surface was protected against air humidity. After some time of exposure, a constant tensile load was applied and held fixed until cracking was observed. The fracture surfaces were then carefully examined and the locations of crack initiation and propagation were identified. This ingenious testing protocol showed that brittle cracking occurred, with cracks initiating from the outer surface, near the notch tip and far away from the corrosive environment. Since hydrogen diffuses through the crystal lattice and accumulates near stress concentrators such as notches or cracks, these findings revealed the important role of hydrogen embrittlement in the SCC of Al alloys.\\

Mimicking the experiments by \citet{Gruhl1984} (see also \citealp{Ratke1980}), we model a cylindrical specimen subjected to constant uniaxial stress with the dimensions given in Fig. \ref{fig:Expmodel2}b. The experimental specimens have five identical surface notches, each separated by a distance of 30 mm. We find through preliminary simulations that the notch interaction effects are negligible and thus choose to simulate a single notch in a region of $\text{H}=30$ mm height. In terms of material properties, the mechanical behaviour of the heat treated AlZn5Mg3 is characterised by a Young's modulus of $E=71.7$ GPa, a Poisson's ratio of $\nu=0.33$, a yield strength of $\sigma_y=550$ MPa, and a strain hardening exponent of $N=0.2$ \citep{DiegoFuentes2018}. From the fracture toughness measurements by \citet{DiegoFuentes2018}, the critical energy release rate in an inert environment is estimated to be $G_c\left(0\right) = 8.5$ kJ/m$^2$. The hydrogen diffusion coefficient is taken to be $D_\mathrm{H}=9.3\times10^{-6}$ mm$^2$/s, as reported by \citet{Young2003}. \citet{Gruhl1984} and \citet{Ratke1980} found that cracking was of intergranular nature, and thus we consider a grain boundary trap binding energy in (\ref{eq:theta}). Specifically, a value of $\Delta g_b^0=19.3$ kJ/mol is adopted (see \citealp{Su2019}). The hydrogen damage coefficient is taken to be equal to $\chi=0.67$, as estimated for Al alloys in the first principles calculations conducted by \citet{Jiang2004a}. Regarding the electrochemical parameters, we follow \citet{Nguyen2018} and assume a metal ion diffusion coefficient of $D_\mathrm{M}=5.41\times10^{-4}$ mm$^2$/s, an interface energy of $\gamma=120$ $\text{J}/\text{m}^2$, a free energy density curvature of $A=733$ N/mm$^2$, and a normalised saturated concentration of $c_\mathrm{Le}=0.03$. The interface thickness in this case study is chosen to be $\ell_d=0.03$ mm, and the interface kinetics coefficient is initially defined as $L_0=1\times10^{-9}$ $\mathrm{mm^2/(N \cdot s)}$, but this magnitude is varied to investigate its influence. The fracture length scale parameter is defined to be $\ell_f=0.1$ mm, as estimated from (\ref{eq:sigma_c}) and a tensile strength of $\sigma_c=825$ MPa. Also, we adopt $\beta_e=1$ and $\beta_p=$0.1 in (\ref{eq:Weightingfactors}). As in the experiments, the sample is subjected to a constant load, resulting in a remote applied stress of $\sigma_a=200$ MPa, which is applied after 24 hours of exposure to the corrosive environment. In our model, the vertical displacements are constrained so as to obtain an average remote stress of 200 MPa. Taking advantage of symmetry, we employ axisymmetric finite elements and accordingly modify the strain-displacement matrix. Also, only half of the 2D section is simulated, with appropriate symmetry boundary conditions along the symmetry line. The finite element mesh is refined in the expected pitting and SCC regions; after a sensitivity study, we employ approximately 20,000 quadratic quadrilateral elements with reduced integration. We emphasise that the outer surface is protected and not exposed to an aggressive environment and thus the penalty condition $c_\text{H}=c_\text{env}$ is only defined at the inner surface.\\

\begin{figure}[H]
\centering
\noindent\makebox[\textwidth]{%
\includegraphics[scale=0.48]{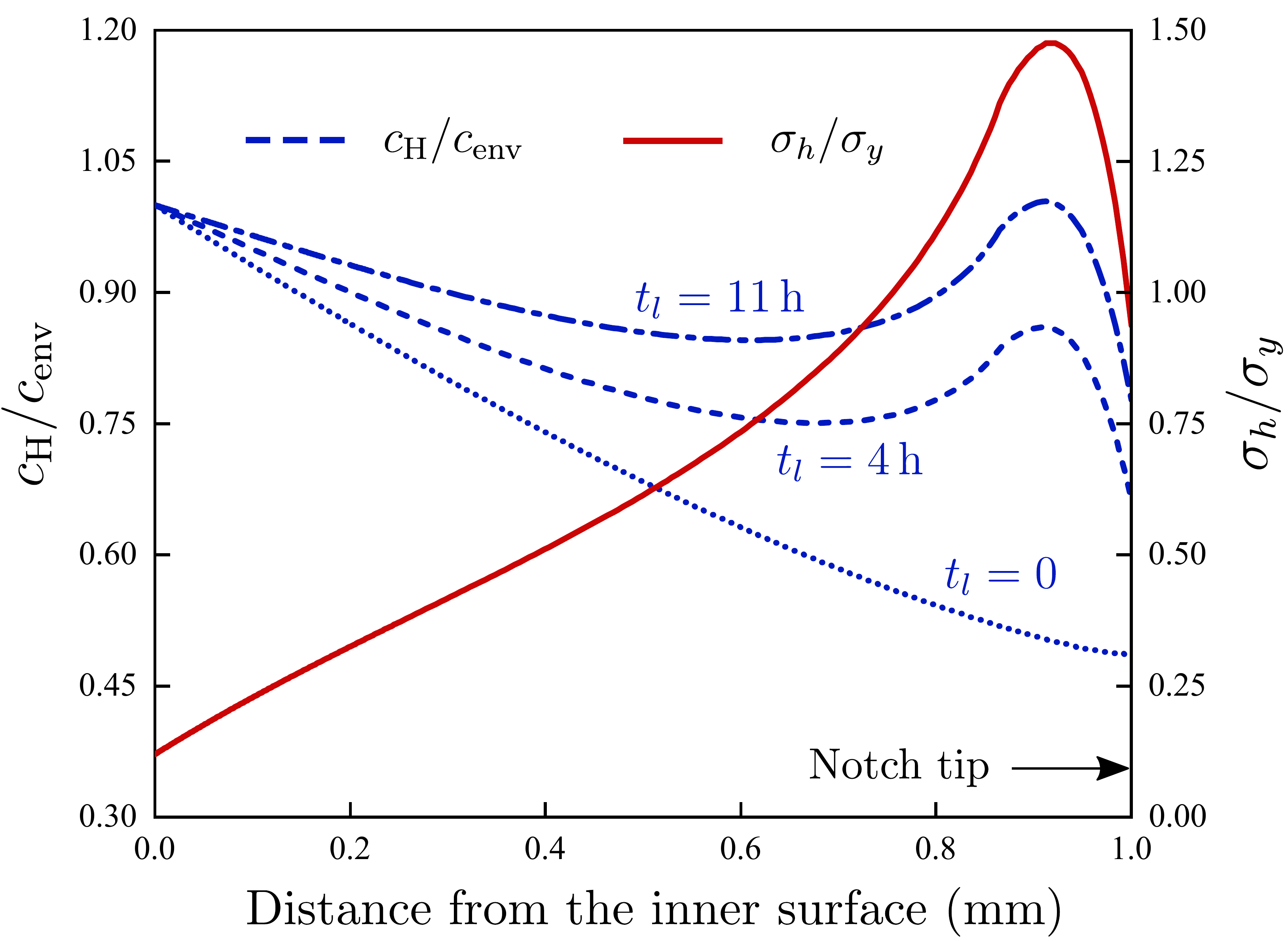}}
\caption{\citet{Gruhl1984} experiments: normalised hydrogen concentration and hydrostatic stress distributions along the sample in the absence of phase field corrosion and damage ($\phi_d=\phi_f=1 \,\, \forall \,\, \mathbf{x}$). Here, $t_l$ denotes the time that has passed since a constant load is applied (i.e., after a pre-exposure time of 24 h).}
\label{fig:H content}
\end{figure}

First, simulations are conducted in the absence of corrosion or damage (i.e., $\phi_d=\phi_f=1 \,\, \forall \,\, \mathbf{x}$) to gain insight into the diffusion of hydrogen within the sample. The experimental protocol involved an initial charging time (with no load) followed by a period combining environmental exposure and application of a constant tensile stress. For a stress of 200 MPa, failure was found to occur after 11 h, with cracks initiating solely near the tip of the notch. It was thus speculated that SCC cracking in AlZn5Mg3 was the result of the transport of hydrogen from the corrosive medium to the vicinity of the notch. The results obtained in this work in terms of stress and hydrogen distribution appear to strengthen that hypothesis. As shown in Fig. \ref{fig:H content}, hydrogen starts accumulating near the notch tip, at a distance from the inner surface that coincides with the peak stress, reaching a magnitude comparable to that of the inner surface after 11 h of loading time. Thus, it appears sensible to deduce that a critical combination of stress and hydrogen content has been achieved after 11 h, which has led to hydrogen-driven SCC from the notch tip. It should be noted that the consideration of plastic strain gradients would have displaced the stress and hydrogen peaks even closer to the notch tip (see \citealp{TAFM2017}).

\begin{figure}[H]
\centering
\noindent\makebox[\textwidth]{%
\includegraphics[scale=0.25]{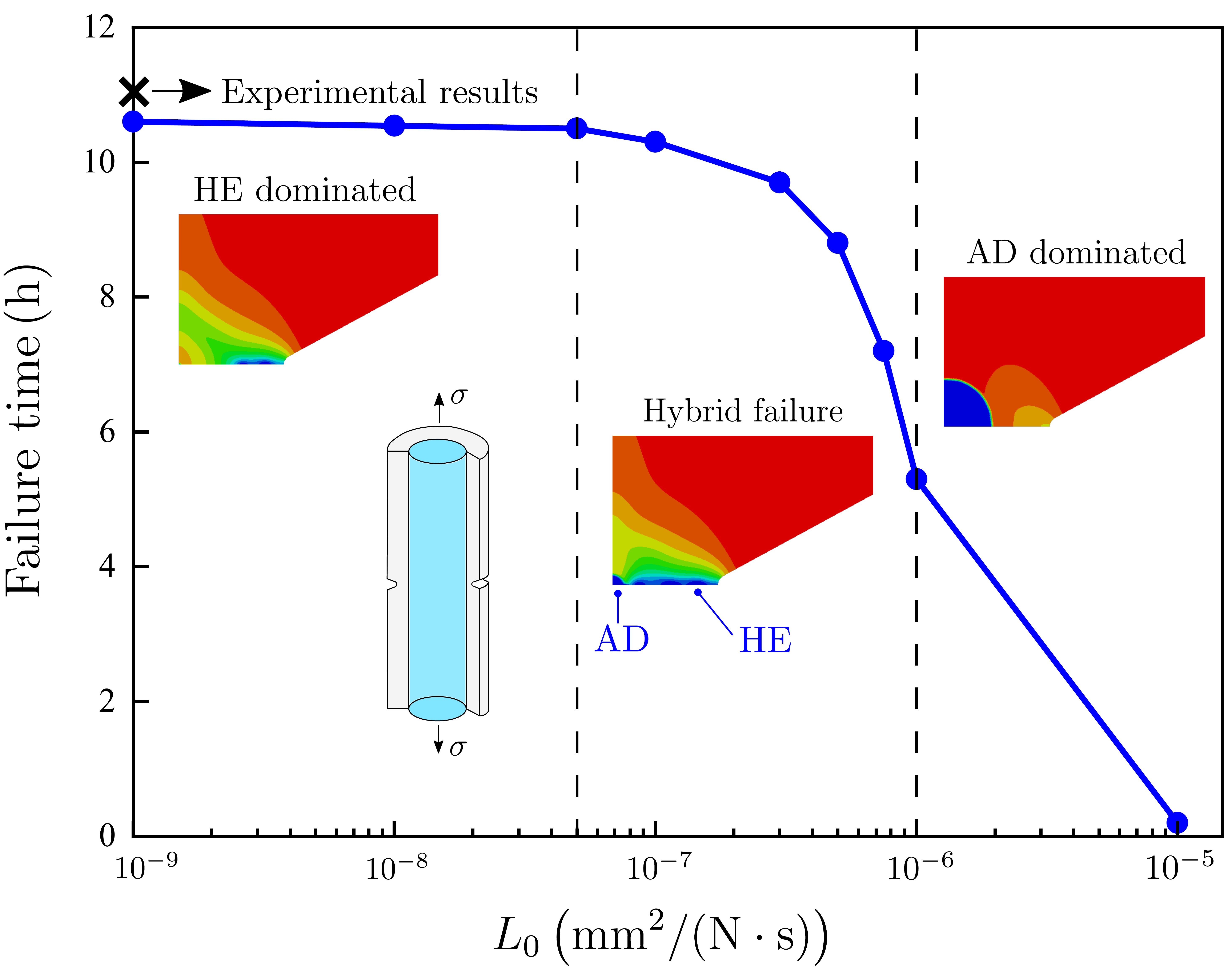}}
\caption{\citet{Gruhl1984} experiments: failure time versus interface kinetics coefficient $L_0$. Phase field $\phi_e$ contours are embedded to showcase the competing failure mechanisms and their dependence of $L_0$.  Increasing $L_0$ is associated with higher corrosion current densities and thus a transition is observed between hydrogen embrittlement, hybrid anodic dissolution and hydrogen embrittlement, and anodic dissolution.}
\label{fig:Failuretime2}
\end{figure}

We proceed to gain further insight by explicitly simulating the corrosion and cracking processes. To this end, a small pit of radius $\text{R}=0.02$ mm is introduced in the inner surface to trigger anodic dissolution. The magnitude of $c_\text{env}$ was not reported and can range from tens to hundreds of parts per million in weight \citep{Scully2012}. A value of $c_\text{env}=72$ wppm is adopted, as this leads to a good approximation to the failure times reported by \citet{Gruhl1984}. As shown at the left edge of Fig. \ref{fig:Failuretime2}, a good agreement is attained with the experimental reported failure times for these assumptions. As can be observed in the embedded contours, failure is driven by hydrogen damage, with cracks nucleating close to the tip of the notch and propagating towards the inner sample surface. 

Then, the interface mobility coefficient $L_0$ is varied to investigate the dependency of the failure mechanism on the environment. As elaborated in Section \ref{Sec:pit-to-crack transition}, $L_0$ is proportional to the corrosion current density $i$ and thus increasing its value allows us to investigate more aggressive corrosive environments. As shown in Fig. \ref{fig:Failuretime2}, increasing $L_0$ leads to a shift in the failure mechanism and a drop in the failure time. Initially, the failure time appears to be insensitive to $L_0$, as failure is purely driven by hydrogen embrittlement and as result the failure time is governed by the diffusion process; 10.6 h are needed for enough hydrogen to accumulate at the tip of the notch. However, as $L_0$ increases beyond 10$^{-7}$ mm$^2$/(N$\cdot$s), pitting is observed to occur, which lowers the overall stiffness of the sample and triggers earlier failures. In this regime, the failure mechanism is a combination of hydrogen embrittlement and anodic dissolution as pitting initiates first but the main crack extends from the notch pit. Dissolution becomes dominant for $L_0$ values larger than 10$^{-6}$ mm$^2$/(N$\cdot$s), where pitting corrosion becomes significant even during the pre-charging time, leading to an almost sudden drop in the load carrying capacity upon the application of a remote displacement. The results shown in Fig. \ref{fig:Failuretime2} illustrate the competition between the kinetics of anodic- and hydrogen-driven failure mechanisms, and extend the seminal experiments by \citet{Gruhl1984} into regimes of higher corrosion currents, where anodic dissolution could have played a dominant role. 

\section{Discussion and summary}
\label{Sec:Conclusions}

We have presented a generalised theory for stress corrosion cracking (SCC), incorporating both hydrogen embrittlement and material dissolution-driven damage mechanisms. The main ingredients of our formulation are: (i) a Cahn-Hilliard type evolution law for the mass-conserved diffusion of metal ions in the aqueous electrolyte; (ii) a KKS-based phase field model for describing material dissolution; (iii) an extended version of Fick's law to characterise the diffusion of hydrogen atoms within the metallic lattice; (iv) a power-law elastic-plastic constitutive material model; and (v) a Griffith-based phase field description of the solid-crack interface. These elements are strongly coupled to capture key phenomena such as the interplay between diffusion and activation controlled corrosion, the role of film rupture and repassivation, the enhancement of corrosion kinetics due to mechanical contributions, the accumulation of hydrogen in areas of high volumetric strains, the sensitivity of material toughness to hydrogen content, and the synergy of multiple mechanisms in driving material degradation.\\

A numerical framework was also presented, which takes as degrees-of-freedom the displacement components, the phase field corrosion order parameter, the normalised concentration of metallic ions, the hydrogen concentration, and the phase field fracture order parameter. The resulting finite element model was used to conduct numerical experiments of particular relevance. First, the model was verified by reproducing two paradigmatic benchmarks in, respectively, corrosion and hydrogen embrittlement; the growth of a semi-circular pit and the failure of a notched square plate exposed to a hydrogen-containing environment. Then, insight was gained into the interaction and competition between the kinetics and damage mechanisms of hydrogen embrittlement and anodic dissolution. Firstly, SCC under pure bending in stainless steel was investigated, showcasing the important role of film passivation in preventing hydrogen uptake and in governing the transition from anodic dissolution driven SCC growth to hydrogen assisted fracture. Secondly, the threshold stress intensity factor for SCC was quantified for a wide range of applied potentials and two different environments: sterile and biologically active seawater. The results showed a very good agreement with experiments and showcased the capability of the model in capturing the change in damage mechanism observed when the applied potential is varied. Finally, we have provided the first simulation results for the experiments by \citet{Gruhl1984} on Al alloys. The finite element results obtained provide a mechanistic interpretation of these seminal experiments and extend the findings to more aggressive corrosive environments, quantifying the threshold at which hydrogen embrittlement ceases to dominate SCC in Al alloys. Through these numerical experiments, new physical insight is gained. In particular, it is worth emphasising and discussing the following findings:
\begin{itemize}
    \item The interplay between mechanics and material dissolution accelerates the localisation of corrosion. As pits and other SCC defects become sharper, this leads to a localised rupture of the passive film and higher dissolution rates due to higher local mechanical fields ($\sigma_h$, $\varepsilon^p$).
    \item Defects that initially grow due to localised anodic dissolution can be driven by hydrogen embrittlement mechanisms at later stages, due to the increased accumulation of hydrogen with time and the rise in hydrostatic stress (and thus hydrogen content) that results for the sharpening of the SCC defect.
   \item The drop in toughness observed for applied potentials larger than the corrosion potential can be both due to increasing dissolution rates rates with $E_p$ or due to the larger generation of hydrogen at the crack faces (or a combination of both).
    \item While hydrogen degradation laws based on first principles are appealing, as they require no calibration, fully predictive estimations are hindered by the uncertainties associated with the estimation of trap binding energies. On the other hand, quantitative estimates of material dissolution can be attained based on the corrosion current density.
    \item Film repassivation and hydrogen exhibit a complex interplay. A more stable film results in a reduced hydrogen uptake but also requires failure loads of higher magnitude, which favour a transition from anodic dissolution to hydrogen embrittlement.
\end{itemize}

The model presented not only enables investigating the interplay between hydrogen and anodic dissolution-driven damage mechanisms but also provides a generalised platform for predicting SCC across for arbitrary choices of material, loading conditions and environment. 

\section*{Acknowledgements}
\label{Sec:Acknowledgeoffunding}

C. Cui acknowledges valuable discussions with Dr Z. Cui and Dr H. Tian  from the Ocean University of China. C. Cui and R. Ma acknowledge financial support from the National Natural Science Foundation of China (grants 52178153 and 51878493). E. Mart\'{\i}nez-Pa\~neda acknowledges financial support from the EPSRC [grant EP/V009680/1] and from UKRI's Future Leaders Fellowship programme [grant MR/V024124/1]. C. Cui additionally acknowledges financial support from the China Scholarship Council (grant 202006260917).



\bibliographystyle{elsarticle-harv}
\bibliography{library}
\end{document}